\documentclass[journal,10pt]{IEEEtran}

\usepackage[font=scriptsize,caption=false,labelsep=space]{subfig}
\usepackage{mathrsfs}
\usepackage{graphicx,float,wrapfig,epstopdf,amsmath}
\usepackage[]{algorithmicx}
\usepackage{algpseudocode,algorithm}
\usepackage{balance}

\usepackage{amsmath}
\usepackage{amssymb} 
\usepackage{amsthm}
\usepackage{mathtools}

\usepackage{graphicx}
\usepackage{epstopdf}
\usepackage{times}
\usepackage{textcomp,cite}
\usepackage{color}
\usepackage{url}
\usepackage{cuted}

\usepackage{multirow}
\usepackage[table]{xcolor}
\usepackage{threeparttable}

\DeclareMathOperator*{\maximize}{maximize} 
\DeclareMathOperator*{\minimize}{minimize} 
\DeclareMathOperator*{\subjectto}{subject \hspace{3pt} to:\hspace{3pt}} 

\begin{document}
	
	\title{A Family of Deep Learning Architectures for Channel Estimation and Hybrid Beamforming in Multi-Carrier mm-Wave Massive MIMO}
	\author{Ahmet~M.~Elbir,~\IEEEmembership{Senior Member,~IEEE,} Kumar Vijay~Mishra,~\IEEEmembership{Senior Member,~IEEE,} M.~R.~Bhavani~Shankar,~\IEEEmembership{Senior Member,~IEEE,} and~Bj\"{o}rn~Ottersten,~\IEEEmembership{Fellow,~IEEE}
		\thanks{This work was supported in part by the  ERC Project  AGNOSTIC.}
		\thanks{A. M. E. is with the Department of Electrical and Electronics Engineering, Duzce University, Duzce, Turkey, and with University of Luxembourg, Luxembourg City, Luxembourg (e-mail: ahmetmelbir@gmail.com).}
		\thanks{K. V. M., M. R. B. S., and B. O. are with the Interdisciplinary Centre for Security, Reliability and Trust (SnT), University of Luxembourg, Luxembourg City L-1855, Luxembourg. E-mail: \{kumar-mishra@ext, bhavani.shankar@, bjorn.ottersten@\}uni.lu.}
	}
	\maketitle
	
	\begin{abstract}
		
		Hybrid analog and digital beamforming transceivers are instrumental in addressing the challenge of expensive hardware and high training overheads in the next generation millimeter-wave (mm-Wave) massive MIMO (multiple-input multiple-output) systems. However, lack of fully digital beamforming in hybrid architectures and short coherence times at mm-Wave impose additional constraints on the channel estimation. Prior works on addressing these challenges have focused largely on narrowband channels wherein optimization-based or greedy algorithms were employed to derive hybrid beamformers. In this paper, we introduce a deep learning (DL) approach for  channel estimation and hybrid beamforming for frequency-selective, wideband mm-Wave systems. In particular, we consider a massive MIMO Orthogonal Frequency Division Multiplexing (MIMO-OFDM) system and propose three different DL frameworks comprising convolutional neural networks (CNNs), which accept the raw data of received signal as input and yield channel estimates and the hybrid beamformers at the output. We also introduce both offline and online prediction schemes. Numerical experiments demonstrate that, compared to the current state-of-the-art optimization and DL methods, our approach provides higher spectral efficiency, lesser computational cost and fewer number of pilot signals, and higher tolerance against the deviations in the received pilot data, corrupted channel matrix, and propagation environment.
	\end{abstract}
	\begin{IEEEkeywords}
		Channel estimation, deep learning, online learning, hybrid beamforming, mm-Wave.
	\end{IEEEkeywords}

	\section{Introduction}
	\label{sec:Introduciton}
	The conventional cellular communications systems suffer from spectrum shortage while the demand for wider bandwidth 
	and higher data rates is continuously increasing \cite{mimoOverview}. In this context, millimeter wave (mm-Wave) band is a preferred candidate for fifth-generation (5G) communications technology because it provides higher data rate and wider bandwidth \cite{mimoOverview,mishra2019toward,5GwhatWillItBe,hodge2019reconfigurable,ayyar2019robust}. Compared to sub-6 GHz transmissions envisaged in 5G, the mm-Wave signals encounter a more complex propagation environment that is characterized by higher scattering, severe penetration losses, lower diffraction, and higher path loss for fixed transmitter and receiver gains \cite{mimoHybridLeus2}. The mm-Wave systems leverage massive antenna arrays - usually in a multiple-input multiple-output (MIMO) configuration - to achieve array and multiplexing gain, and thereby compensate for the propagation losses at high frequencies \cite{mimoRHeath,liu2020codesign}. 
	
	However, such a large array requires a dedicated radio-frequency (RF) chain for each antenna resulting in an expensive system architecture and high power consumption. In order to address this, hybrid analog and baseband beamforming architectures have been introduced, wherein a small number of phase-only analog beamformers are employed to steer the beams. The down-converted signal is then processed by baseband beamformers, each of which is dedicated to a single RF chain \cite{mimoHybridLeus2,mimoRHeath,mimoScalingUp}.
	This combination of high-dimensional phase-only analog and low-dimensional baseband digital beamformers significantly  reduces the number of RF chains while also maintaining sufficient beamforming gain \cite{mmwaveKeyElements,mimoRHeath}. 	However, lack of fully digital beamforming in hybrid architectures poses challenges in mm-Wave channel estimation \cite{channelEstLargeArrays,channelEstLargeArrays2,channelEstimation1,channelEstimation1CS}.
	
	The instantaneous channel state information (CSI) is essential for massive MIMO communications because precoding at the transmitted or decoding at the receiver transmission requires highly accurate CSI to achieve spatial diversity and multiplexing gain \cite{mimoHybridLeus2}. In practice, pilot signals are periodically transmitted and the received signals are processed to estimate the CSI \cite{channelEstLargeArrays,channelEstLargeArrays2}. 
	Further, the mm-Wave environments such as indoor and vehicular communications are highly variable with short coherence times \cite{coherenceTimeRef} that necessitates use of channel estimation algorithms that are robust to deviations in the channel data. Once the CSI is obtained, the hybrid analog and baseband beamformers are designed \cite{mimoHybridLeus3,hybridBFAltMin,hybridBFLowRes,sohrabiOFDM}. 
	
	In recent years, several techniques have been proposed to design the hybrid precoders in mm-Wave MIMO systems. Initial works have focused on narrow-band channels \cite{mimoHybridLeus2,mimoHybridLeus3,mimoRHeath,hybridBFLowRes}. However, to effectively utilize the mm-Wave MIMO architectures with relatively larger bandwidth, there are recent and concerted efforts toward developing broadband hybrid beamforming techniques. The key challenge in hybrid beamforming for a broadband frequency-selective channel is designing a common analog beamformer that is shared across all subcarriers while the digital (baseband) beamformer weights need to be specific to a subcarrier. This difference in  hybrid beamforming design of frequency-selective channels from flat-fading case is the primary motivation for considering hybrid beamforming for orthogonal frequency division multiplexing (OFDM) modulation. The optimal beamforming vector in a frequency-selective channel depends on the frequency, i.e., a subcarrier in OFDM, but the analog beamformer in any of the narrow-band hybrid structures cannot vary with frequency. Thus, a common analog beamformer must be designed in consideration of impact to all subcarriers, thereby making the hybrid precoding more difficult in the frequency-selective case than in the narrow-band case. \textcolor{black}{While it is possible to align generated beams at different subcarriers to the same physical direction of the mobile user via frequency-to-angle mapping, these techniques require additional time-delayer network, which increases the hardware complexity~\cite{thz_wb_tracking}.  }
	
	Among prior works, \cite{widebandChannelEst1,widebandChannelEst2,channelEstWBHeath1,channelEstWBHeath2} consider channel estimation for wideband mm-Wave massive MIMO systems. The hybrid beamforming design was investigated in \cite{alkhateeb2016frequencySelective,sohrabiOFDM,widebandHBWithoutInsFeedback,widebandMLbased} where OFDM-based frequency-selective structures are designed. In particular, \cite{alkhateeb2016frequencySelective} proposes a Gram-Schmidt orthogonalization based approach for hybrid beamforming (GS-HB) with the assumption of perfect CSI and GS-HB selects the precoders from a finite codebook which are obtained from the instantaneous channel data. Using the same assumption on CSI, \cite{sohrabiOFDM} proposed a phase extraction approach for hybrid precoder design. In \cite{zhu2016novel}, a unified analog beamformer is designed based on the second-order spatial channel covariance matrix of a wideband channel. 
	
	Nearly all of the aforementioned methods strongly rely on perfect CSI knowledge. This is very impractical given the highly dynamic nature of mm-Wave channel \cite{coherenceTimeRef}. 	To relax this dependence and obtain robust performance against the imperfections in the estimated channel matrix, we examine a deep learning (DL) approach.	The DL is capable of uncovering complex relationships in data/signals and, thus, can achieve better performance in terms of spectral efficiency. This has been demonstrated in several successful applications of DL in wireless communications problems such as channel estimation \cite{deepCNN_ChannelEstimation,dl_CE1}, analog beam selection \cite{mimoDLHybrid,hodge2019multi}, and also hybrid beamforming \cite{mimoDLHybrid,mimoDeepPrecoderDesign,elbirQuantizedCNN2019,elbirHybrid_multiuser}. In particular, DL-based techniques have been shown  \cite{deepCNN_ChannelEstimation,deepLearningCommOverAir,elbirIETRSN2019,elbirQuantizedCNN2019} to be computationally efficient in searching for optimum beamformers and tolerant to imperfect channel inputs when compared with the conventional methods. {\color{black} For instance in \cite{directHB_Alkhateeb}, beamformer vectors are design based on the compressive channel data which is learned via deep auto-encoders.} However, these works investigated only narrow-band channels \cite{mimoDeepPrecoderDesign,elbirQuantizedCNN2019}, including our previous work in \cite{elbirQuantizedCNN2019}. {\color{black}DL-based channel estimation is studied in~\cite{deepCNN_ChannelEstimation} for broadband mm-Wave signals. However, this approach requires massive training of the channel and the input data of the DL network are obtained via the computationally expensive step of matrix inversion.} {\color{black}The DL-based wideband beamforming in ~\cite{mimoDLChannelModelBeamformingFacebook} considers multiple BSs that collect the omni-beampatterns from the mobile user for codebook-based beamforming. The performance of this approach is dependent on the accuracy and the resolution of the beamformers in the codebook while optimum solution can be achieved~\cite{elbirQuantizedCNN2019,hybridBFAltMin}. } {\color{black}Furthermore, DL-based statistical hybrid beamforming is investigated in~\cite{elbirTVTWithoutCSI}, which assumes the knowledge of perfect channel covariance matrix at the transmitter.}
	
	
	In this paper, we introduce \textcolor{black}{a set of} DL architectures for channel estimation and hybrid beamformer in wideband mm-Wave systems. Preliminary results of this work appeared in \cite{elbir2020low} where a single deep network was used for only wideband hybrid beamforming. In this paper, our proposed frameworks {\color{black}extend this idea to both channel estimation and hybrid beamformer design in both offline and online learning schemes by} 
	constructing a non-linear mapping between the received pilot signals and the channel data or hybrid beamformers. The online models allow us to estimate the channel and the beamformers when the channel statistics are changed. We employ convolutional neural networks (CNNs) in three different DL structures. In the first framework (F1), a single CNN maps the received pilot signals directly to the hybrid beamformers. In the second (F2) and third (F3) frameworks, we employ multiple CNNs to also estimate the channel and {\color{black}beamformers separately. In F2, entire subcarrier data are jointly fed to a single CNN for channel estimation.} This is a less complex architecture but it does not allow flexibility of controlling each channel individually. Therefore, we modify the structure of F2 in F3, which has a dedicated CNN for each subcarrier. {\color{black}Each CNN is fed with the raw data of the received signals. This allows us to estimate the channel via DL using much fewer number of pilot
		signals, which leads to a significant reduction in the channel training procedures. Thus, the purpose of the CNN is to learn the hidden features inheriting in the raw data and construct a mapping to the desired data, i.e., the channel and beamformers.} {\color{black}The main contributions of this paper are as follows:
		\begin{enumerate}
			\item We introduce a set of DL architectures for channel estimation and hybrid beamforming in wideband  mm-Wave systems. While the prior works include DL-based beamformer design, they either assume the availability of perfect CSI~\cite{mimoDLChannelModelBeamformingFacebook} or operate in narrowband scenario~\cite{elbirQuantizedCNN2019,mimoDeepPrecoderDesign} even though it is more relevant to investigate the mm-Wave channel as a frequency-selective channel. 
			\item Compared to the prior DL-based channel estimation techniques~\cite{deepCNN_ChannelEstimation}, the proposed channel estimation method needs approximately $3$ times fewer pilot signals to construct the full channel matrix and exhibits satisfactory spectral efficiency performance.
			\item Nearly all of the prior DL-based works consider offline learning for channel estimation or hybrid beamforming~\cite{mimoDLHybrid,mimoDeepPrecoderDesign,deepLearningCommOverAir,directHB_Alkhateeb,mimoDLChannelModelBeamformingFacebook}. This paper introduces an online method, wherein the learning model infrequently updates its parameters in accordance with the changes in the input data. 
	\end{enumerate}}
	
	The proposed DL framework operates in two stages: offline training and online prediction. During training, to obtain {\color{black}the labels of the learning model,} several received pilot signals and channel realizations are generated, and the hybrid beamforming problem is solved via the manifold optimization (MO) approach \cite{manopt}.  In the online prediction stage when the CNNs operate in real-time, the channel matrix and the hybrid beamformers are estimated by simply feeding the CNNs with the received pilot data.	The proposed approach is advantageous because it does not require perfect channel data in the prediction stage yet it provides robust performance. {\color{black}Compared to the conventional approaches, our DL approach has superior performance in terms of robustness as well as less computational time {\color{black}and fewer pilot signals} to produce hybrid beamformers. In addition, the model-free architecture of the proposed DL-based approach is a promising way to reduce power consumption and hardware complexity, for future artificial-intelligence-based communication systems.}

	\begin{figure*}[t]
		\centering
		{\includegraphics[width=.7\textwidth]{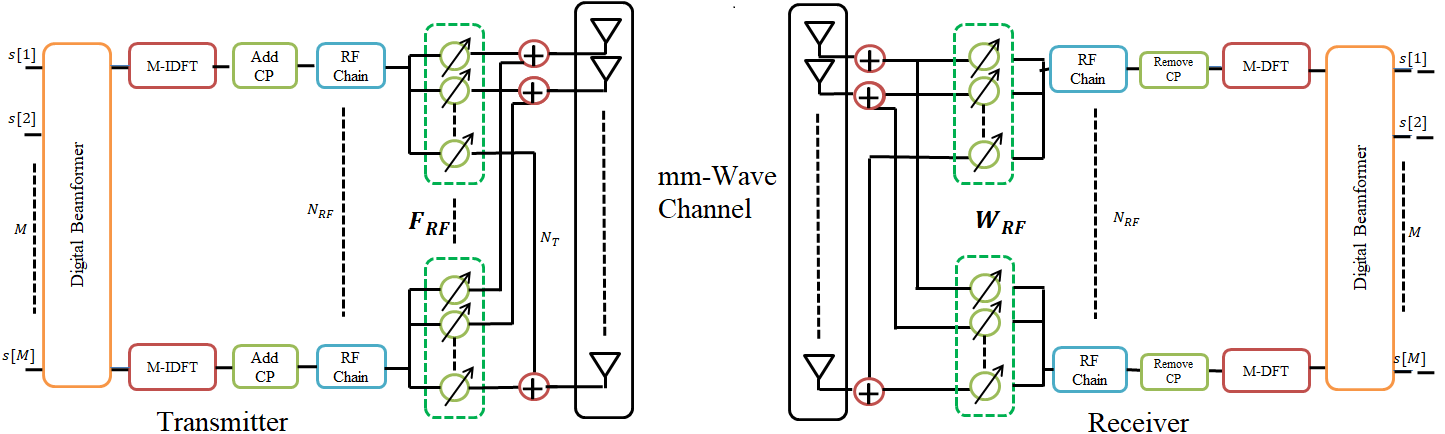} }
		\caption{System architecture of wideband mm-Wave MIMO-OFDM transceivers with hybrid analog and baseband precoding. }
		\label{fig_SystemArchitecture}
	\end{figure*}
	
	Furthermore, we have investigated the online deployment of the proposed DL frameworks. The main challenge in online learning is to adapt the network to the propagation environment so that it can learn the new incoming input data. In the proposed scheme, the DL network updates its parameters so that it adapts to the environment. Since the proposed network architectures work in supervised manner, the label of the networks are obtained by solving analytical approaches only when the network requires to be re-trained. 
	
	
	Throughout this paper, we denote the vectors and matrices by boldface lower and upper case symbols, respectively. In case of a vector $\mathbf{a}$, $[\mathbf{a}]_{i}$ represents its $i$-th element. For a matrix $\mathbf{A}$, $[\mathbf{A}]_{:,i}$ and $[\mathbf{A}]_{i,j}$ denote the $i$-th column and the $(i,j)$-th entry, respectively. The $\mathbf{I}_N$ is the identity matrix of size $N\times N$; $\mathbb{E}\{\cdot\}$ denotes the statistical expectation; $\textrm{rank}(\cdot)$ denotes the rank of its matrix argument; $\|\cdot\|_\mathcal{F}$ is the Frobenius norm; $(\cdot)^{\dagger}$ denotes the Moore-Penrose pseudo-inverse; and $\angle\{\cdot\}$ denotes the angle of a complex scalar/vector. The notation expressing a convolutional layer with $N$ filters/channels of size $D\times D$, is given by  $N$@$ D\times D$.

	\section{System Model}
	\label{sec:SystemModel}
	We consider the hybrid precoder design for a frequency selective wideband mm-Wave massive MIMO-OFDM system with $M$ subcarriers (Fig.~\ref{fig_SystemArchitecture}). The base station (BS) has $N_\mathrm{T}$ antennas and $N_\mathrm{RF}$ $(N_\mathrm{RF} \leq N_\mathrm{T})$ RF chains to transmit $N_\mathrm{S}$ data streams. In the downlink, the BS first precodes $N_\mathrm{S}$ data symbols $\mathbf{s}[m] = [s_1[m],s_2[m],\dots,s_{N_\mathrm{S}}[m]]^\textsf{T}\in \mathbb{C}^{N_\mathrm{S}}$ at each subcarrier by applying the subcarrier-dependent baseband precoders $\mathbf{F}_{\mathrm{BB}}[m] = [\mathbf{f}_{\mathrm{BB}_1}[m],\mathbf{f}_{\mathrm{BB}_2}[m],\dots,\mathbf{f}_{\mathrm{BB}_{N_\mathrm{S}}} [m]]\in \mathbb{C}^{N_{\mathrm{RF}}\times N_\mathrm{S}}$. {\color{black}Then, as shown in  Fig.~1, the signal is transformed to the time-domain via $M$-point inverse fast Fourier transform (IFFT)}. After adding the cyclic prefix, the transmitter employs a subcarrier-independent RF precoder $\mathbf{F}_{\mathrm{RF}}\in \mathbb{C}^{N_\mathrm{T}\times N_{\mathrm{RF}}}$ to form the transmitted signal. Given that $\mathbf{F}_{\mathrm{RF}}$ consists of analog phase shifters, we assume that the RF precoder has constant equal-norm elements, i.e., $|[\mathbf{F}_{\mathrm{RF}}]_{i,j}|^2 =1$. Additionally, we have the power constraint  $\sum_{m=1}^{M}\|\mathbf{F}_{\mathrm{RF}}\mathbf{F}_{\mathrm{BB}}[m] \|_\mathcal{F}^2= MN_\mathrm{S}$ that is enforced by the normalization of baseband precoder $\{\mathbf{F}_{\mathrm{BB}}[m] \}_{m\in \mathcal{M}}$ where $\mathcal{M} = \{1,\dots,M\}$. Thus, the $N_\mathrm{T}\times 1$ transmit signal is
	\begin{align}
	\mathbf{x}[m] = \mathbf{F}_{\mathrm{RF}} \mathbf{F}_{\mathrm{BB}}[m]  \mathbf{s}[m].
	\end{align}
	In mm-Wave transmission, the channel is represented by a geometric model with limited scattering \cite{mimoChannelModel1}. The channel matrix $\mathbf{H}[m]$ includes the contributions of $L$ clusters, each of which has the time delay $\tau_l$ and $N_\mathrm{sc} $ scattering paths/rays within the cluster. Hence, each ray ($r = \{1,\dots, N_\mathrm{sc}\}$) in the $l$-th cluster has a relative time delay $\tau_{{r}}$, angle-of-arrival (AOA) $\theta_l \in [-\pi,\pi]$, angle-of-departure (AOD) $\phi_l \in [-\pi,\pi]$, relative AOA (AOD) shift $\vartheta_{rl}$ ($\varphi_{rl}$) between the center of the cluster and each ray \cite{alkhateeb2016frequencySelective},  and complex path gain $\alpha_{l,r}$. Let $p(\tau)$ denote a pulse shaping function for $T_\mathrm{s}$-spaced signaling evaluated  at $\tau$ seconds \cite{channelModelSparseSayeed}, then the mm-Wave delay-$d$ MIMO channel matrix is
	\begin{align}
	\label{eq:delaydChannelModel}
	\bar{\mathbf{H}}[d] =  \sqrt{\frac{ N_\mathrm{T} N_{\mathrm{R}} } {N_{sc}L}}\sum_{l=1}^{L} \sum_{r=1}^{N_\mathrm{sc}}\alpha_{l,r} p(dT_\mathrm{s} - \tau_l - \tau_{{r}})
	\nonumber \\ \times \mathbf{a}_\mathrm{R}(\theta_{l} - \vartheta_{rl}) \mathbf{a}_\mathrm{T}^\textsf{H}(\phi_l - \varphi_{rl}),
	\end{align}
	where $\mathbf{a}_\mathrm{R}(\theta)$ and $\mathbf{a}_\mathrm{T}(\phi)$ are the $N_\mathrm{R} \times 1$ and $N_\mathrm{T}\times 1$ steering vectors representing the array responses of the receive and transmit antenna arrays respectively. 
	Let $\lambda_m = \frac{c_0}{f_m}$ be the wavelength for the subcarrier $m$ with frequency of $f_m$. Since the operating frequency is relatively higher than the bandwidth in mm-Wave systems and the subcarrier frequencies are close to each other, (i.e., $f_{m_1} \approx f_{m_2}$, $m_1,m_2 \in\mathcal{M}$), we use a single operating wavelength for the analog beamformers as $\lambda = \lambda_{1} = \dots = \lambda_{M} = \frac{c_0}{f_c}$ where $c_0$ is speed of light and $f_c$ is the central carrier frequency \cite{sohrabiOFDM}. This approximation also allows for a single frequency-independent analog beamformer for each subcarrier.  Then, for a uniform linear array (ULA), the array response of the transmit array is
	\begin{align}
	\mathbf{a}_\mathrm{T}(\phi) = \big[ 1, e^{j\frac{2\pi}{\lambda} \overline{d}_\mathrm{T}\sin(\phi)},\dots,e^{j\frac{2\pi}{\lambda} (N_\mathrm{T}-1)\overline{d}_\mathrm{T}\sin(\phi)} \big]^\textsf{T},
	\end{align}
	where $\overline{d}_\mathrm{T}=\overline{d}_\mathrm{R} = \lambda/2$ is the antenna spacing and  $\mathbf{a}_\mathrm{R}(\theta)$ can be defined in a similar way as for $\mathbf{a}_\mathrm{T}(\phi)$. {\color{black} After performing $M$-point DFT of the delay-$d$ channel model in (\ref{eq:delaydChannelModel}), the channel matrix at subcarrier $m$ is
		\begin{align}
		\label{Hm_OFDM}
		\mathbf{H}[m] = \sum_{d=0}^{D-1}\bar{\mathbf{H}}[d]e^{-j\frac{2\pi m}{M} d},
		\end{align}
		where $D$ is the length of cyclic prefix \cite{channelModelSparseBajwa}. The frequency domain channel in (\ref{Hm_OFDM}) is used in MIMO-OFDM systems, where orthogonality is satisfied as $\mathbb{E}\{||\mathbf{H}^\textsf{H}[m_1]\mathbf{H}[m_2]||_\mathcal{F}^2\} = 0 $ for $m_1, m_2 \in \mathcal{M}$ and $m_1 \neq m_2$. We also note here that the application of precoder and combiners does not violate orthogonality since the baseband precoder/combiners are obtained for each subcarrier and analog precoder/combiners are the same for all subcarriers. }

	{\color{black}With the aforementioned block-fading channel model, the received signal before analog processing at subcarrier $m$ is
		\begin{align}
		\label{arrayOutput}
		\mathbf{y}[m] = \sqrt{\rho}\mathbf{H}[m] \mathbf{F}_\mathrm{RF}\mathbf{F}_\mathrm{BB}[m]\mathbf{s}[m] + \mathbf{n}[m],
		\end{align}
		where $\rho$ represents the average received power and  $\mathbf{H}[m]\in \mathbb{C}^{N_\mathrm{R}\times N_\mathrm{T}}$ channel matrix and $\mathbf{n}[m] \sim \mathcal{CN}(\mathbf{0},\sigma^2 \mathbf{I}_\mathrm{N_\mathrm{R}})$ is additive white Gaussian noise (AWGN) vector. The received signal is first processed by the analog combiner $\mathbf{W}_\mathrm{RF}$. We assume that $N_\mathrm{RF}$ RF chains are employed at the receiver, and the cyclic prefix is removed from the processed signal and $N_\mathrm{RF}$ $M$-point FFTs are applied to yield the signal in frequency domain (see Fig.~\ref{fig_SystemArchitecture}). Finally, the receiver employs low-dimensional $N_\mathrm{RF}\times N_\mathrm{S}$ digital combiners $\{\mathbf{W}_\mathrm{BB}[m]\}_{m\in \mathcal{M}}$. The received and processed signal is $\widetilde{\mathbf{y}}[m] = \mathbf{W}_\mathrm{BB}^\textsf{H}[m]\mathbf{W}_\mathrm{RF}^\textsf{H}\mathbf{y}[m]$, i.e., 
		\begin{align}
		\label{sigModelReceived}
		\widetilde{\mathbf{y}}[m] =  \sqrt{\rho}\mathbf{W}_\mathrm{BB}^\textsf{H}[m]\mathbf{W}_\mathrm{RF}^\textsf{H}\mathbf{H}[m] \mathbf{F}_\mathrm{RF}\mathbf{F}_\mathrm{BB}[m]\mathbf{s}[m] 
		\nonumber \\ + \mathbf{W}_\mathrm{BB}^\textsf{H}[m]\mathbf{W}_\mathrm{RF}^\textsf{H}\mathbf{n}[m],
		\end{align}
		where the analog combiner $\mathbf{W}_\mathrm{RF}\in \mathbb{C}^{N_\mathrm{R}\times N_\mathrm{RF}}$ has the constraint $\big[[\mathbf{W}_\mathrm{RF}]_{:,i}[\mathbf{W}_\mathrm{RF}]_{:,i}^\textsf{H}\big]_{i,i}=1$ similar to the RF precoder. }
	
	{\color{black}Once the received symbols, i.e., $	\widetilde{\mathbf{y}}[m]$ are obtained at the receiver, they are demodulated and decoded for each subcarrier. In this work, the OFDM structure is fully loaded, i.e., all subcarriers are activated.}
	
	\section{Problem Formulation}
	\label{sec:probform}
	We focus on designing hybrid precoders $\mathbf{F}_\mathrm{RF},\mathbf{F}_\mathrm{BB}[m]$, $\mathbf{W}_\mathrm{RF},\mathbf{W}_\mathrm{BB}[m]$ by maximizing the overall spectral efficiency of the system under {\color{black}total power spectral density constraint of all subcarriers}. Let $R[m]$ be the overall spectral efficiency of the subcarrier $m$. Assuming that the Gaussian symbols are transmitted through the mm-Wave channel \cite{mimoRHeath,mimoHybridLeus2,alkhateeb2016frequencySelective}, $R[m]$ is given by
	\begin{align}
	R[m] =  \textrm{log}_2 \bigg| \mathbf{I}_{N_\mathrm{S}} 
	+\frac{\rho}{N_\mathrm{S}}\boldsymbol{\Lambda}_\mathrm{n}^{-1}[m]\mathbf{W}_\mathrm{BB}^\textsf{H}\mathbf{W}_\mathrm{RF}^\textsf{H} 
	\nonumber \\ \times  \mathbf{H}[m]\mathbf{{F}}_\mathrm{RF}\mathbf{{F}}_\mathrm{BB}[m]\mathbf{{F}}_\mathrm{BB}^\textsf{H}[m] \mathbf{{F}}_\mathrm{RF}^\textsf{H}\mathbf{H}^\textsf{H}[m]\mathbf{W}_\mathrm{RF}\mathbf{W}_\mathrm{BB}[m] \bigg|,\nonumber
	\end{align}
	where $\boldsymbol{\Lambda}_\mathrm{n}[m] = \sigma_n^2 \mathbf{W}_\mathrm{BB}^\textsf{H}[m]\mathbf{W}_\mathrm{RF}^\textsf{H} \mathbf{W}_\mathrm{RF}\mathbf{W}_\mathrm{BB}[m]\in \mathbb{C}^{N_\mathrm{S} \times N_\mathrm{S}}$ corresponds to the noise term in (\ref{sigModelReceived}). The hybrid beamformer design is equivalent to the following optimization problem:
	\begin{align}
	\label{HBdesignProblem}
	&\underset{\mathbf{{F}}_\mathrm{RF},\mathbf{{W}}_\mathrm{RF}, \{\mathbf{{F}}_\mathrm{BB}[m],\mathbf{{W}}_\mathrm{BB}[m]\}_{m\in \mathcal{M}}}{\maximize} \frac{1}{M}\sum_{m =1}^{M} R[m] \nonumber \\
	&\subjectto \; \mathbf{{F}}_\mathrm{RF} \in \mathcal{F}_\mathrm{RF},   \mathbf{{W}}_\mathrm{RF} \in \mathcal{W}_\mathrm{RF},
	\nonumber \\& \sum_{m=1}^{M}||\mathbf{{F}}_\mathrm{RF}\mathbf{{F}}_\mathrm{BB}[m]||_{\mathcal{F}}^2 = M N_\mathrm{S},
	\end{align}
	where $\mathcal{F}_\mathrm{RF}$ and $\mathcal{W}_\mathrm{RF}$ are the feasible sets for the RF precoder and combiners which obey the unit-norm constraint.

	The hybrid beamformer design problem in (\ref{HBdesignProblem}) requires analog and digital beamformers which, in turn, are obtained by exploiting the structure of the channel matrix in mm-Wave channel. 
	Our goal is to recover the mm-Wave channel $\mathbf{H}[m]$ for the given received pilot signal. In the following section, we first describe the channel estimation and design methodology of hybrid beamformers before introducing learning-based approach. 
	
	In practice, the estimation process of the channel matrix is a challenging task, especially in the case of a large number of antennas deployed in massive MIMO communications \cite{channelEstLargeArrays,channelEstimation1}. Further, short coherence times of mm-Wave channel imply that the channel characteristics change rapidly \cite{coherenceTimeRef}. Literature indicates several mm-Wave channel estimation techniques \cite{mimoChannelModel2,channelEstimation1CS,channelEstimation1,mimoAngleDomainFaiFai,mimoHybridLeus2}. 
	In our DL framework, the channel estimation is performed by a deep network which accepts the received pilot signals as input and yields the channel matrix estimate at the output layer~\cite{deepCNN_ChannelEstimation}. During the pilot transmission process, the transmitter activates only one RF chain to transmit the pilot on a single beam; the receiver meanwhile turns on all RF chains \cite{mimoHybridLeus2}. Hence, unlike other DL-based beamformers \cite{elbirQuantizedCNN2019,mimoDeepPrecoderDesign} that presume knowledge of the channel, our framework exploits DL for both channel matrix approximation as well as beamforming. 

	{\color{black}In the following, we first address how the  input and output data of the DL networks are obtained for  both channel estimation and hybrid beamforming problems respectively, then introduce the proposed DL frameworks.
		
	}
	\section{Channel Estimation} 
	\label{sec:ice}
	{In this work, DL network estimates the channel from the received pilot signals in the preamble stage. Consider a downlink scenario where the transmitter activates only one RF chain. Let us define the beamformer vector as $\overline{\mathbf{f}}_u[m]\in\mathbb{C}^{N_\mathrm{T}}$ and pilot signals are $\overline{{s}}_u[m]$, where  $u = 1,\dots,M_\mathrm{T}$. {\color{black}The receiver employs  a total of  $M_\mathrm{R}$ ($M_\mathrm{R}\leq N_\mathrm{R}$) combining vectors and applies $\overline{\mathbf{w}}_v$ for $v = 1,\dots, M_\mathrm{R}$ to process the received pilots \cite{deepCNN_ChannelEstimation}. Since the number of RF chains in the receiver is limited by $N_\mathrm{RF}$ (usually less than $M_\mathrm{R}$ in a single channel use), only $N_\mathrm{RF}$ combining vectors can be employed  in a single channel use.} Hence, the total channel use in the channel acquisition process is $\lceil \frac{M_\mathrm{R}}{N_\mathrm{RF}}\rceil$.
		
		{\color{black}
			After processing through combiners, the received pilot signal takes the form
			\begin{align}
			\label{receivedSignalPilot}
			\mathbf{{Y}}[m] = \overline{\mathbf{W}}^\textsf{H}[m] \mathbf{H}[m] \overline{\mathbf{F}}[m]\overline{\mathbf{S}}[m] + \widetilde{\mathbf{N}}[m],
			\end{align}
			where $\overline{\mathbf{F}}[m] = [\overline{\mathbf{f}}_1[m],\overline{\mathbf{f}}_2[m],\dots,\overline{\mathbf{f}}_{M_\mathrm{T}}[m]]$ and $\overline{\mathbf{W}}[m] = [\overline{\mathbf{w}}_1[m],\overline{\mathbf{w}}_2[m],\dots,\overline{\mathbf{w}}_{M_\mathrm{R}}[m]]$ are $N_\mathrm{T}\times M_\mathrm{T}$ and $N_\mathrm{R}\times M_\mathrm{R}$ beamformer matrices.} The $\overline{\mathbf{S}}[m] = \mathrm{diag}\{ \overline{s}_1[m],\dots,\overline{s}_{M_\mathrm{T}}[m]\}$ denotes pilot signals and $\widetilde{\mathbf{N}}[m]= \overline{\mathbf{W}}^\textsf{H} \overline{\mathbf{N}}[m]$ is effective noise matrix, where $\overline{\mathbf{N}}[m] \sim \mathcal{N}(0, \sigma_{\overline{\mathbf{N}}}^2)$. The noise corruption of the pilot training data is measured by SNR$_{\overline{\mathbf{N}}}$. {Without loss of generality, we assume that $\overline{\mathbf{F}}[m] = \overline{\mathbf{F}}\in \mathbb{C}^{N_\mathrm{T}\times M_\mathrm{T}}$ and $\overline{\mathbf{W}}[m] = \overline{\mathbf{W}}\in \mathbb{C}^{N_\mathrm{R}\times M_\mathrm{R}}$, $\forall m$ so that the beamformers are not subcarrier-dependent in the pilot training~\cite{mimoAngleDomainFaiFai}. A common approach is that $\overline{\mathbf{F}}[m]$ and $\overline{\mathbf{W}}[m]$ are selected as the first $M_\mathrm{T}$ columns of an $N_\mathrm{T}\times N_\mathrm{T}$ discrete Fourier transform (DFT) matrix and the first $M_\mathrm{R}$ columns of an $N_\mathrm{R}\times N_\mathrm{R}$ DFT matrix respectively~\cite{deepCNN_ChannelEstimation}. {\color{black}Also, we set $\overline{\mathbf{S}}[m] = \sqrt{P_\mathrm{T}}\mathbf{I}_{M_\mathrm{T}}$, where $P_\mathrm{T}$ is the transmit power because pilot design is not the main focus of this work. However, there are several works on the design  of the training pilots in one symbol/frame for OFDM structures, see e.g.,~\cite{pilotDesign_OFDM}}.} Then, the received signal in (\ref{receivedSignalPilot}) becomes
		\begin{align}
		\label{receivedSignalPilotMod}
		\mathbf{{Y}}[m] = \overline{\mathbf{W}}^\textsf{H} \mathbf{H}[m] \overline{\mathbf{F}} + \widetilde{\mathbf{N}}[m].
		\end{align}
		{\color{black}Here, $\mathbf{Y}[m]\in\mathbb{C}^{M_\mathrm{R}\times M_\mathrm{T}}$ is selected as input to feed the DL network.  In contrast, previous works, such as~\cite{deepCNN_ChannelEstimation}, involve a preprocessing stage to  obtain a tentative channel estimate as $(\overline{\mathbf{W}}\overline{\mathbf{W}}^\textsf{H})^{-1}\overline{\mathbf{W}}\mathbf{{Y}}[m]\overline{\mathbf{F}}^\textsf{H}(\overline{\mathbf{F}}\overline{\mathbf{F}}^\textsf{H})^{-1}$, which requires $M_\mathrm{T}\geq N_\mathrm{T}$ and $M_\mathrm{R} \geq N_\mathrm{R}$. \textcolor{black}{Compared to~\cite{deepCNN_ChannelEstimation}, the proposed method is   advantageous because it does not require a matrix inversion, which is computationally expensive in practice. Furthermore, the proposed method does not require $M_\mathrm{T}\geq N_\mathrm{T}$ and $M_\mathrm{R} \geq N_\mathrm{R}$, which allows us to estimate the channel data accurately with much less degrees of freedom in the prediction stage by exploiting DL methods that do not rely on the statistics of the data. Instead, DL only constructs a non-linear mapping between the input ($\mathbf{Y}[m]\in\mathbb{C}^{M_\mathrm{R}\times M_\mathrm{T}}$) and output ($\mathbf{H}[m]\in\mathbb{C}^{N_\mathrm{R}\times N_\mathrm{T}}$) data~\cite{directHB_Alkhateeb} (see, e.g., Fig.~\ref{fig_CE_Mt})}. 
		}

	}

	\section{Hybrid Beamformer Design For Wideband mm-Wave MIMO Systems}
	\label{sec:bb_hb}
	
	The design problem in (\ref{HBdesignProblem}) requires a joint optimization over several matrices. This approach is computationally complex and even intractable. Instead, a decoupled problem is preferred{\color{black}, where the precoders are designed first by assuming that the receiver performs an optimal decoding. While decoupling provides an efficient way to transceiver design, the  price paid for this approach is the absence of the optimal receiver architecture assumed during precoder design. Nevertheless, revisiting the combiner design problem and use the available information by the estimated precoders provide a realistic design strategy~\cite{mimoRHeath,sohrabiOFDM}.} Here, the hybrid precoders $\mathbf{F}_\mathrm{RF},\mathbf{F}_\mathrm{BB}[m]$ are estimated first and then the hybrid combiners $\mathbf{W}_\mathrm{RF},\mathbf{W}_\mathrm{BB}[m]$ are found. 
	Define the mutual information of the mm-Wave channel that can be achieved at the BS through Gaussian signaling as~\cite{alkhateeb2016frequencySelective}
	\begin{align}
	\mathcal{I}\{\mathbf{F}_\mathrm{RF},\mathbf{F}_\mathrm{BB}[m]\} = 
	\nonumber \\ \textrm{log}_2 \bigg| \mathbf{I}_{N_\mathrm{S}}  +\frac{\rho}{N_\mathrm{S}}\mathbf{H}[m]\mathbf{{F}}_\mathrm{RF}\mathbf{{F}}_\mathrm{BB}[m]\mathbf{{F}}_\mathrm{BB}^\textsf{H}[m] \mathbf{{F}}_\mathrm{RF}^\textsf{H}\mathbf{H}^\textsf{H}[m] \bigg|.
	\end{align}
	The hybrid precoders are then obtained by maximizing the mutual information, i.e.,
	\begin{align}
	\label{PrecoderDesignProblem}
	&\underset{\mathbf{{F}}_\mathrm{RF}, \{\mathbf{{F}}_\mathrm{BB}[m]\}_{m\in \mathcal{M}}}{\maximize} \frac{1}{M}\sum_{m =1}^{M} \mathcal{I}\{\mathbf{F}_\mathrm{RF},\mathbf{F}_\mathrm{BB}[m]\} \nonumber \\
	&\subjectto \; \mathbf{{F}}_\mathrm{RF} \in \mathcal{F}_\mathrm{RF}, \sum_{m=1}^{M}||\mathbf{{F}}_\mathrm{RF}\mathbf{{F}}_\mathrm{BB}[m]||_{\mathcal{F}}^2 = M N_\mathrm{S},
	\end{align}
	
	The optimization problem in (\ref{PrecoderDesignProblem}) can be approximated by exploiting the similarity between the hybrid beamformer $\mathbf{F}_\mathrm{RF}\mathbf{F}_\mathrm{BB}[m]$ and the optimal unconstrained beamformer $\mathbf{F}^{\mathrm{opt}}[m]$. The latter is obtained from the right singular matrix of the channel matrix $\mathbf{H}[m]$ \cite{hybridBFAltMin,mimoRHeath}. Let the singular value decomposition of the channel matrix be  $\mathbf{H}[m] = \mathbf{U}[m] \boldsymbol{\Sigma}[m] \mathbf{V}^H[m]$, where $\mathbf{U}[m]\in \mathbb{C}^{N_\mathrm{R}\times \mathrm{rank}(\mathbf{H}[m])}$ and $\mathbf{V}[m]\in \mathbb{C}^{N_\mathrm{T} \times \mathrm{rank}(\mathbf{H}[m])}$ are the left and the right singular value matrices of the channel matrix, respectively, 
	and $\boldsymbol{\Sigma}[m]$ is $\mathrm{rank}(\mathbf{H}[m])\times \mathrm{rank}(\mathbf{H}[m])$ matrix composed of the singular values of $\mathbf{H}[m]$ in descending order. 
	
	By decomposing $\boldsymbol{\Sigma}[m]$ and $\mathbf{V}[m]$ as $\boldsymbol{\Sigma}[m] = \mathrm{diag}\{ \widetilde{\boldsymbol{\Sigma}}[m],\overline{\boldsymbol{\Sigma}}[m] \},\hspace{5pt} \mathbf{V}[m] = [\widetilde{\mathbf{V}}[m],\overline{\mathbf{V}}[m]],$
	where $\widetilde{\mathbf{V}}[m]\in \mathbb{C}^{N_\mathrm{T}\times N_\mathrm{S}}$, the unconstrained precoder is readily obtained as $\mathbf{F}^{\mathrm{opt}}[m] = \widetilde{\mathbf{V}}[m]$ \cite{mimoRHeath}. The hybrid precoder design problem for all subcarriers can then be written as 
	\begin{align}
	\label{PrecoderAllCarriers}
	&\underset{\mathbf{F}_\mathrm{RF},\{\mathbf{F}_\mathrm{BB}[m]\}_{m \in \mathcal{M}}}{\minimize} \big|\big|   \widetilde{\mathbf{F}}^{\mathrm{opt}}  - \mathbf{F}_\mathrm{RF}\widetilde{\mathbf{F}}_\mathrm{BB}  \big|\big|_\mathcal{F}^2
	\nonumber \\
	&\subjectto \; \mathbf{{F}}_\mathrm{RF} \in \mathcal{F}_\mathrm{RF}, \sum_{m=1}^{M}\big|\big| \mathbf{{F}}_\mathrm{RF}\mathbf{{F}}_\mathrm{BB}[m]\big|\big|_{\mathcal{F}}^2 =  MN_\mathrm{S},
	\end{align}
	where  $ \widetilde{\mathbf{F}}^{\mathrm{opt}}\in \mathbb{C}^{N_\mathrm{T}\times MN_\mathrm{S}}$ is defined as 
	\begin{align}
	\widetilde{\mathbf{F}}^{\mathrm{opt}} = \begin{bmatrix} \mathbf{F}^{\mathrm{opt}}[1]
	& \mathbf{F}^{\mathrm{opt}}[2] & \cdots & \mathbf{F}^{\mathrm{opt}}[M] \end{bmatrix},
	\end{align}
	and 
	\begin{align}
	\widetilde{\mathbf{F}}_\mathrm{BB} = \begin{bmatrix} \mathbf{{F}}_\mathrm{BB}[1] & \mathbf{{F}}_\mathrm{BB}[2] & \cdots & \mathbf{{F}}_\mathrm{BB}[M] \end{bmatrix},
	\end{align}
	contain the beamformers for all subcarriers.
	%

	Once the hybrid precoders are designed, the hybrid combiners $\mathbf{W}_\mathrm{RF},\mathbf{W}_\mathrm{BB}[m]$ realized by minimizing the mean-square-error (MSE), $\mathbb{E}\{\big|\big| \mathbf{s}[m] - \mathbf{W}_\mathrm{BB}^\textsf{H}[m] \mathbf{W}_\mathrm{RF}^\textsf{H}\mathbf{y}[m]  \big|\big|_2^2\}$. Using  the minimum MSE (MMSE) estimator $\mathbf{W}_\mathrm{MMSE}[m]= (\mathbb{E}\{\mathbf{s}[m] \mathbf{y}^\textsf{H}[m]  \} \mathbb{E}\{\mathbf{y}[m]  \mathbf{y}^\textsf{H}[m]   \}^{-1})^\textsf{H}$, we can write the combiner design problem  as 
	%
	\begin{align}
	\label{CombinerOnlyProblemEquivalent}
	&\underset{\mathbf{W}_\mathrm{RF}, \mathbf{W}_\mathrm{BB}[m]}{\minimize}
	\big|\big| \boldsymbol{\Lambda}_\mathrm{y}^{1/2}[m] (\mathbf{W}_\mathrm{MMSE}[m] 
	- \mathbf{W}_\mathrm{RF} \mathbf{W}_\mathrm{BB}[m] )\big|\big|_\mathcal{F}^2 \nonumber \\
	&\subjectto \mathbf{W}_\mathrm{RF} \in{\mathcal{W}}_\mathrm{RF}.
	\end{align}
	where $\boldsymbol{\Lambda}_\mathrm{y}[m] = 
	\rho\mathbf{H}[m]\mathbf{F}_\mathrm{RF}\mathbf{F}_\mathrm{BB}[m]\mathbf{F}_\mathrm{BB}^\textsf{H}[m]\mathbf{F}_\mathrm{RF}^\textsf{H}\mathbf{H}^\textsf{H}[m] + \sigma_n^2\mathbf{I}_{N_\mathrm{R}},$
	is the covariance of the array output in (\ref{arrayOutput}). 
	The unconstrained combiner in a compact form is then,
	\begin{align}
	\mathbf{W}_\mathrm{MMSE}^\textsf{H}[m] = \frac{1}{\rho}\bigg( \mathbf{F}^{\mathrm{opt}^\textsf{H}}[m]\mathbf{H}^\textsf{H}[m]\mathbf{H}[m]\mathbf{F}^{\mathrm{opt}}[m] \nonumber \\
	+ \frac{N_\mathrm{S}\sigma_n^2}{\rho}\mathbf{I}_{N_\mathrm{S}} \bigg)^{-1}  \mathbf{F}^{\mathrm{opt}^\textsf{H}}[m]\mathbf{H}^\textsf{H}[m].
	\end{align}
	In (\ref{CombinerOnlyProblemEquivalent}), the multiplicative term $\boldsymbol{\Lambda}_\mathrm{y}^{1/2}[m]$ does not depend on $\mathbf{W}_\mathrm{RF}$ or $\mathbf{W}_\mathrm{BB}[m]$. It, therefore, has no bearing on the solution and can be ignored. 
	Define $\widetilde{\mathbf{W}}_\mathrm{MMSE}\in \mathbb{C}^{N_\mathrm{R}\times MN_\mathrm{S}}$ as
	\begin{align}
	\widetilde{\mathbf{W}}_\mathrm{MMSE} \hspace{-3pt} =\hspace{-3pt} \big[\mathbf{W}_\mathrm{MMSE}[1] \hspace{3pt} {\mathbf{W}}_\mathrm{MMSE}[2]\cdots{\mathbf{W}}_\mathrm{MMSE}[M]\big],
	\end{align}
	and 
	\begin{align}
	\widetilde{\mathbf{W}}_\mathrm{BB} = \begin{bmatrix}{\mathbf{W}}_\mathrm{BB}[1] & {\mathbf{W}}_\mathrm{BB}[2] & \cdots &{\mathbf{W}}_\mathrm{BB}[M]  \end{bmatrix}.
	\end{align}
	Then, the hybrid combiner design problem becomes
	\begin{align}
	\label{CombinerOnlyProblemAllSubcarriers}
	&\underset{\mathbf{W}_\mathrm{RF}, \{\mathbf{W}_\mathrm{BB}[m]\}_{m\in \mathcal{M}}}{\minimize}
	\big|\big| \widetilde{\mathbf{W}}_\mathrm{MMSE}
	- \mathbf{W}_\mathrm{RF} \widetilde{\mathbf{W}}_\mathrm{BB}\big|\big|_\mathcal{F}^2 \nonumber \\
	&\subjectto \;\;\;\;\;\mathbf{W}_\mathrm{RF} \in{\mathcal{W}}_\mathrm{RF} \nonumber \\
	& \mathbf{W}_\mathrm{BB}[m] = (\mathbf{W}_\mathrm{RF}^\textsf{H} \boldsymbol{\Lambda}_\mathrm{y}[m]  \mathbf{W}_\mathrm{RF})^{-1} (\mathbf{W}_\mathrm{RF}^\textsf{H}\boldsymbol{\Lambda}_\mathrm{y}[m]\mathbf{W}_\mathrm{MMSE}[m]).
	\end{align}
	In \cite{manopt}, manifold optimization or ``Manopt'' algorithm is suggested to effectively solve the optimization problems in (\ref{PrecoderAllCarriers}) and  (\ref{CombinerOnlyProblemAllSubcarriers}). {\color{black}Although the Manopt algorithm is computationally complex as compared to the conventional greedy search~\cite{mimoRHeath,alkhateeb2016frequencySelective} and phase extraction-based~\cite{sohrabiOFDM} approach, it yields optimal beamforming results which are used as labels of deep networks. The optimality of MO is in the sense that it achieves the minimum Euclidean distance between the unconstrained and hybrid beamformers. This complexity influences computational times of only offline training. When the network is deployed for online testing, the proposed DL approach avoids this computational overhead without running the Manopt algorithm for new inputs.} Note that both of these problems do not require a codebook or a set of array response of transmit and receive arrays \cite{mimoRHeath}. In fact, the manifold optimization problem for (\ref{PrecoderAllCarriers}) and (\ref{CombinerOnlyProblemAllSubcarriers}) are initialized at a random point, i.e., beamformers with unit-norm and random phases.

	\section{Learning-Based Channel Estimation and Hybrid Beamformer Design}
	\label{sec:HD_Design}
	\begin{figure}[t]
		\centering
		\includegraphics[draft=false,width=1\columnwidth]{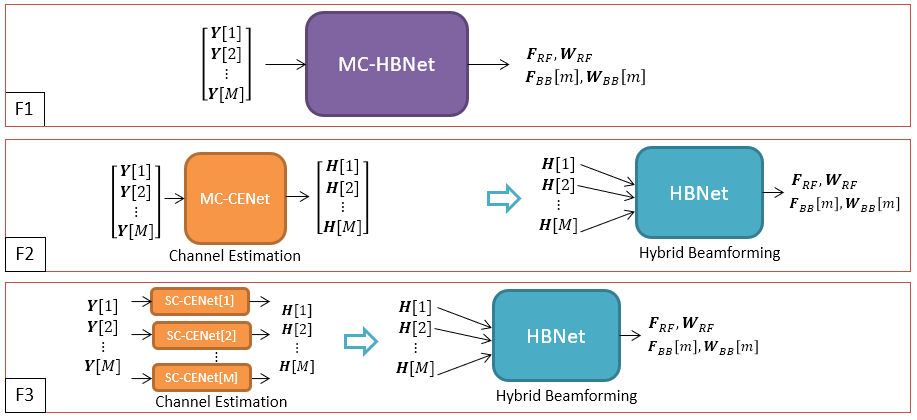} 
		\caption{\color{black}Deep learning frameworks for hybrid beamforming in mm-Wave MIMO systems. The F1 has a single CNN (MC-HBNet) which maps the received pilot signal data directly into hybrid beamformers. In F2 and F3, multiple CNNs are used for channel estimation and hybrid beamforming sequentially. For channel estimation, a single CNN (MC-CENet) is trained for all subcarrier data in F2 whereas a dedicated CNN (SC-CENet) is used for each subcarrier data in F3. The final HBNet stage is identical in F2 and F3. }
		\label{fig_DLFrameworks}
	\end{figure}
	
	We introduce three DL frameworks F1, F2, and F3 (Fig.~\ref{fig_DLFrameworks}). In all of them, the received signals $\mathbf{Y}[m]$ are the input and the hybrid beamformers are the outputs. 
	The F1 architecture is Multi-Carrier Hybrid Beamforming Network (MC-HBNet). It comprises a single CNN which accepts the received signals jointly for all subcarriers. The input size is $MM_\mathrm{R} \times M_\mathrm{T}$. {\color{black}The direct use of received signals introduces a performance loss for hybrid beamformer design as compared to the model fed with the channel data.} To address this, F2 employs separate CNNs for channel estimation (Multi-Carrier Channel Estimation Network or MC-CENet) and hybrid beamforming (HBNet). {\color{black}The MC-CENet accepts the received signals of all subcarriers as input, hence, the input size is  $MM_\mathrm{R}\times M_\mathrm{T}$.} To make the setup even more flexible at the cost of computational simplicity, F3 employs one CNN per subcarrier for estimating the channel. For the $m$-th subcarrier, each Single Carrier Channel Estimation Network (SC-CENet$[m]$, $m\in \mathcal{M}$) feeds into a single HBNet. 
	
	\subsection{Input Data}
	We partition the input data into three components to enrich the input features. In our previous works, similar approaches have provided good features for DL implementations \cite{elbirQuantizedCNN2019, elbirIETRSN2019,deepCNN_ChannelEstimation}. In particular, we use the real, imaginary parts and the absolute value of each entry of the input. The absolute value entry indicates to the DL network that the real and imaginary input feeds are connected. {\color{black}The input for MC-HBNet and MC-CENet are the same as the received data of all subcarriers. For instance, the input of F1 is given by $\mathbf{X}_{\mathrm{F1}}  = [\mathbf{X}_{\mathrm{F1}}^\textsf{T}[1],\dots, \mathbf{X}_{\mathrm{F1}}^\textsf{T}[M]  ]^\textsf{T}$. Then, for $M_\mathrm{R}\times M_\mathrm{T}$ received data, the $(i,j)$-th entry of the submatrices per subcarrier is $[[\mathbf{X}_{\mathrm{F1}}[m]]_{:,:,1}]_{i,j} = | [\mathbf{Y}[m]]_{i,j}|$ for the first ``channel'' or input matrix of $\mathbf{X}_{\mathrm{F1}}[m]$.
		The second and the third channels are $[[\mathbf{X}_{\mathrm{F1}}[m]]_{:,:,2}]_{i,j} = \operatorname{Re}\{[\mathbf{Y}[m]]_{i,j}\}$ and $[[\mathbf{X}_{\mathrm{F1}}[m]]_{:,:,3}]_{i,j} = \operatorname{Im}\{[\mathbf{Y}[m]]_{i,j}\}$, respectively. Hence, the size of $\mathbf{X}_{\mathrm{F1}}$ and $\mathbf{X}_{\mathrm{F2}}$ is $M M_\mathrm{R}\times M_\mathrm{T}\times 3$. In F3, the input data for each SC-CENet comprises  a single subcarrier data, hence,  $\mathbf{X}_{\mathrm{F3}}\in \mathbb{R}^{M M_\mathrm{R}\times M_\mathrm{T}\times 3}$. The inputs of HBNet in both F2 and F3  have the same structure; it is denoted as $\mathbf{X}_{\mathbf{H}} = [\mathbf{X}_{\mathbf{H}}^\textsf{T}[1],\dots, \mathbf{X}_{\mathbf{H}}^\textsf{T}[M]  ]^\textsf{T} $ which is of size  $M N_\mathrm{R}\times N_\mathrm{T}\times 3$, where $[[\mathbf{X}_{\mathbf{H}}[m]]_{:,:,1}]_{i,j} = | [\mathbf{H}[m]]_{i,j}|$,  $[[\mathbf{X}_{\mathbf{H}}[m]]_{:,:,2}]_{i,j} = \operatorname{Re}\{[\mathbf{H}[m]]_{i,j}\}$ and $[[\mathbf{X}_{\mathbf{H}}[m]]_{:,:,3}]_{i,j} = \operatorname{Im}\{[\mathbf{H}[m]]_{i,j}\}$.}
	
	\begin{figure}[t]
		\centering
		\includegraphics[draft=false,width=1\columnwidth]{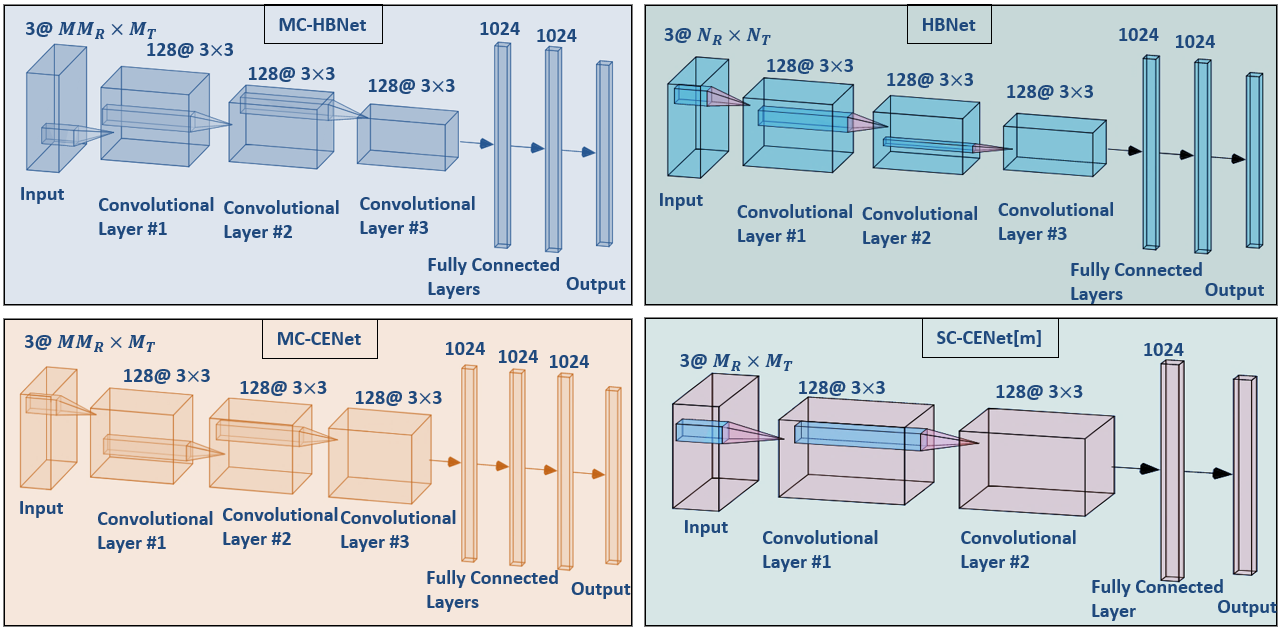} 
		\caption{Deep network architectures used in DL frameworks F1, F2, and F3 for wideband mm-wave channel estimation and hybrid beamforming. }
		\label{fig_Networks}
	\end{figure}
	
	\subsection{Labeling}
	The hybrid beamformers are the common output for all three frameworks (Fig.~\ref{fig_DLFrameworks}). We represent the output as the vectorized form of analog beamformers common to all subcarriers and baseband beamformers corresponding to all subcarriers. The output is an $N_\mathrm{RF}\big(N_\mathrm{T} + N_\mathrm{R} + 2MN_\mathrm{S} \big) \times 1 $ real-valued vector
	\begin{align}
	\label{zSU}
	\hspace{10pt} \mathbf{z} = \begin{bmatrix} \mathbf{z}_\mathrm{RF}^\textsf{T} & \widetilde{\mathbf{z}}_\mathrm{BB}^\textsf{T} \end{bmatrix}^\textsf{T},
	\end{align}
	where $\mathbf{z}_\mathrm{RF} = [\mathrm{vec}\{\angle \mathbf{F}_\mathrm{RF}  \}^\textsf{T},\mathrm{vec}\{\angle \mathbf{W}_\mathrm{RF}  \}^\textsf{T}]^\textsf{T}$ is a real-valued $N_\mathrm{RF}(N_\mathrm{T} + N_\mathrm{R})\times 1$ vector which includes the phases of analog beamformers. The $\widetilde{\mathbf{z}}_\mathrm{BB}\in \mathbb{R}^{2M N_\mathrm{S} N_\mathrm{RF}}$ is composed of the baseband beamformers for all subcarriers as $ \widetilde{\mathbf{z}}_\mathrm{BB} = [\mathbf{z}_\mathrm{BB}^\textsf{T}[1],\mathbf{z}_\mathrm{BB}^\textsf{T}[2],\dots,\mathbf{z}_\mathrm{BB}^\textsf{T}[M]]^\textsf{T} $ where
	\begin{align}
	\mathbf{z}_\mathrm{BB}[m] = [\mathrm{vec}\{\operatorname{Re}\{ \mathbf{F}_\mathrm{BB}[m]\} \}^\textsf{T}, \mathrm{vec}\{\operatorname{Im}\{ \mathbf{F}_\mathrm{BB}[m]\} \}^\textsf{T}, 
	\nonumber  \\\mathrm{vec}\{\operatorname{Re}\{ \mathbf{W}_\mathrm{BB}[m]\} \}^\textsf{T}, \mathrm{vec}\{\operatorname{Im}\{ \mathbf{W}_\mathrm{BB}[m]\} \}^\textsf{T}]^\textsf{T}.\nonumber
	\end{align}

	{\color{black}The output label of the MC-CENet in F2 is the channel matrix of all subcarriers. Given that MC-CENet is fed by  $\mathbf{Y}[m]$, $m\in \mathcal{M}$, the output label for MC-CENet is  $\mathbf{z}_{\mathbf{H}} = [\mathbf{z}_{\mathbf{H}[1]}^\textsf{T},\dots, \mathbf{z}_{\mathbf{H}[M]}^\textsf{T}]^\textsf{T}$, where
		\begin{align}
		\label{zH}
		\mathbf{z}_{\mathbf{H}[m]} = [\mathrm{vec}\{\operatorname{Re}\{\mathbf{H}[m]\}\}^\textsf{T} , \mathrm{vec}\{\operatorname{Im}\{\mathbf{H}[m]\}\}^\textsf{T} ]^\textsf{T},
		\end{align}
	}which is a real-valued vector of size $2N_\mathrm{R}N_\mathrm{T}$. The SC-CENet$[m]$ in F3 has similar input and output structures as the MC-CENet but the received signals $\mathbf{Y}[m]$ are fed to each SC-CENet$[m]$ separately.
	
	
	\subsection{Network Architectures and Offline Training}
	\label{sec:NetTraining}
	
	We design four deep network architectures (Fig.~\ref{fig_Networks}). The MC-HBNet and MC-CENet have input size of $MM_\mathrm{R}\times M_\mathrm{T}\times 3$ whereas the input for  SC-CENet$[m]$ and MC-CENet are  $M_\mathrm{R}\times M_\mathrm{T}\times 3$ and $MM_\mathrm{R}\times M_\mathrm{T}\times 3$, respectively. The number of filters and number of units for all layers are shown in Fig.~\ref{fig_Networks}. There are dropout layers with a $50\%$ probability after each fully connected layer in each network. We use pooling layers after the first and second convolutional layers only in MC-HBNet and HBNet to reduce the dimension by two. The output layer of all networks is the regression layer whose size depends on the application as discussed earlier. 	\textcolor{black}{After a hyperparameter tuning process in which the number of layers and the number of units in each layer are tuned, we fixed the parameters of the learning model, which yields the best performance for the considered scenario \cite{elbirQuantizedCNN2019,elbirIETRSN2019}.}

	\begin{algorithm}
		\begin{algorithmic}[1] \footnotesize
			\caption{Offline training data generation. }
			\Statex {\textbf{Input:} $N$,  $G$, $M$, SNR,  SNR$_{{\mathbf{H}}}$, SNR$_{\overline{\mathbf{N}}}$}. \\
			{\textbf{Output:} Training datasets for the networks in Fig.~\ref{fig_DLFrameworks}: $\mathcal{D}_{\mathrm{MC-HBNet}}$, $\mathcal{D}_{\mathrm{MC-CENet}}$, $\mathcal{D}_{\mathrm{HBNet}}$ and $\mathcal{D}_{\mathrm{SC-CENet}}$.}
			\label{alg:algorithmTraining}
			\State Generate  $\{\mathbf{H}^{(n)}[m]\}_{n=1}^N$  for $m \in \mathcal{M}$.
			\State Initialize with $t=1$ while the dataset length is $T=NG$ for MC-HBNet, HBNet, SC-CENet and MC-CENet.
			\State   \textbf{for}  $1 \leq n \leq N$ \textbf{do}
			\State  \indent \textbf{for}  $1 \leq g \leq G$ \textbf{do}
			\State \indent  $[\widetilde{\mathbf{H}}^{(n,g)}[m]]_{i,j} \sim \mathcal{CN}([\mathbf{H}^{(n)}[m]]_{i,j},\sigma_{\mathbf{H}}^2)$.
			\State \indent {\color{black}Generate $M_\mathrm{R}\times M_\mathrm{T}$} received  signal from (\ref{receivedSignalPilotMod}) as
			\begin{align}
			{\mathbf{Y}}^{(n,g)}[m] = \overline{\mathbf{W}}^{\textsf{H}} \mathbf{H}^{(n,g)}[m] \overline{\mathbf{F}} + \widetilde{\mathbf{N}}^{(n,g)}[m]. \nonumber 
			\end{align}
			\State \indent  Using $\mathbf{H}^{(n,g)}[m]$, find  $\hat{\mathbf{F}}_{\mathrm{RF}}^{(n,g)}$ and $\hat{\mathbf{F}}_{\mathrm{BB}}^{(n,g)}[m]$    by solving  (\ref{PrecoderAllCarriers}).
			\State \indent  Find  $\hat{\mathbf{W}}_{\mathrm{RF}}^{(n,g)}$ and $\hat{\mathbf{W}}_{\mathrm{BB}}^{(n,g)}[m]$   by solving  (\ref{CombinerOnlyProblemAllSubcarriers}).
			\State \indent Input for MC-HBNet and {\color{black}MC-CENet: $\mathbf{X}_{\mathrm{F1}}^{(t)} =[\mathbf{X}_{\mathrm{F1}}^{(t)^\textsf{T}}[1],\dots,\mathbf{X}_{\mathrm{F1}}^{(t)^\textsf{T}}[M] ]^\textsf{T}$  and $\mathbf{X}_{\mathrm{F2}}^{(t)} = \mathbf{X}_{\mathrm{F1}}^{(t)}$,} for $ m\in \mathcal{M}$.
			\State \indent Output for MC-HBNet: $\mathbf{z}_\mathrm{HB}^{(t)} = \mathbf{z}^{(t)}$ as in  (\ref{zSU}).
			\State \indent  {\color{black}Output for MC-CENet: $ \mathbf{z}_{\mathbf{H}}^{(t)}$ as in (\ref{zH}).}
			\State \indent Input for HBNet: $\mathbf{X}_\mathbf{H}^{(t)} = [\mathbf{X}_{\mathbf{H}}^{(t)^\textsf{T}}[1],\dots,\mathbf{X}_{\mathbf{H}}^{(t)^\textsf{T}}[M] ]^\textsf{T}$. Output for HBNet: $\mathbf{z}_\mathrm{HB}^{(t)}$.
			\State \indent  Input for SC-CENet$[m]$: $\mathbf{X}_\mathrm{F3}^{({t})}[m] = \mathbf{X}_{\mathrm{F1}}^{(t)}[m]  $. 
			Output for SC-CENet$[m]$: $\mathbf{z}_{\mathbf{H}[m]}^{({t})}  $. 
			\State \indent $t = t+1$.
			
			\State \indent\textbf{end for} $g$,	
			\State \textbf{end for} $n$,
		\end{algorithmic} 
	\end{algorithm}

	To train the proposed CNN structures, we realize $N=100$ different scenarios for $G=100$ (see Algorithm~\ref{alg:algorithmTraining}) {\color{black} and generate the training datasets for MC-HBNet, MC-CENet, HBNet and SC-CENet as $\mathcal{D}_{\mathrm{MC-HBNet}}$, $\mathcal{D}_{\mathrm{MC-CENet}}$, 
		$\mathcal{D}_{\mathrm{HBNet}} $ and 
		$\mathcal{D}_{\mathrm{SC-CENet}}$ respectively.} For each scenario, we generated a channel matrix and received pilot signal where we introduce additive noise to the training data on both the channel matrix and the received pilot signal which are defined by SNR$_{\mathbf{H}}$ and SNR$_{\overline{\mathbf{N}}}$ respectively\footnote{
		Throughout all numerical experiments, we used four SNR definitions, all of which are characterized by AWGN. 1) SNR$_{\overline{\mathbf{N}}}$ and SNR$_{\overline{\mathbf{N}}-\mathrm{TEST}}$ denote, respectively, training and test stage SNRs corresponding to the signal in (\ref{receivedSignalPilotMod}) when the pilot signals are received in the preamble. 2)  SNR$_{\mathbf{H}}$ and SNR$_{\mathbf{H}-\mathrm{TEST}}$ are, respectively, training and test stage SNRs for the channel matrix used to obtain the corrupted channel data. 3) SNR$_{\overline{\mathbf{S}}-\mathrm{TEST}}$ corresponds to the corruption noise added on the pilot signals. 4) We use	the term ``SNR" on the received signal in (\ref{arrayOutput}) (not in the preamble) for hybrid beamforming process.}. The reason for using different SNRs is to make the DL network robust against fluctuations and imperfections in the data. While training {\color{black} all deep networks}, we use multiple SNR$_{\mathbf{H}}$ and SNR$_{\overline{\mathbf{N}}}$ values to enable robustness in the networks against corrupted input characteristics \cite{elbirQuantizedCNN2019}. In particular, we use  SNR$_{\overline{\mathbf{N}}} = \{20, 30, 40\}$ dB and SNR$_{\mathbf{H}} =\{15,20,25\}$ dB, where we have SNR$_{\mathbf{H}} = 20\log_{10}(\frac{|[\mathbf{H}[m]]_{i,j}|^2}{\sigma_{\mathbf{H}}^2})$ and SNR$_{\overline{\mathbf{N}}} = 20\log_{10}(\frac{|[ \mathbf{H}[m] \overline{\mathbf{F}}[m]\overline{\mathbf{S}}[m]]_{i,j}|^2}{\sigma_{\overline{\mathbf{N}}}^2})$ for $\sigma_{\mathbf{H}}^2$ and $\sigma_{\overline{\mathbf{N}}}^2$ being the corresponding noise variances. In addition, SNR $=\{-10, 0, 10\}$ dB is selected in the training process. As a result, the sizes of the training data for MC-HBNet, MC-CENet, HBNet and SC-CENet$[m]$ are $MM_\mathrm{R}\times M_\mathrm{T}\times 3 \times 30000 $, $MM_\mathrm{R}\times M_\mathrm{T}\times 3 \times 30000 $, $MM_\mathrm{R}\times M_\mathrm{T}\times 3 \times 30000 $ and $N_\mathrm{R}\times N_\mathrm{T}\times 3 \times 30000 $,  respectively. Further, $80\%$ and $20\%$ of all generated data are chosen for training and validation datasets, respectively. For the prediction process, we generated $J_T$ Monte Carlo experiments where a test data which is separately generated by adding noise on received pilot signal with SNR defined as SNR$_{\overline{\mathbf{N}}-\mathrm{TEST}}$. Note that this operation is applied to corrupt input data and test the network against deviations in the input data which can resemble the changes in the channel matrix due to short coherence times in mm-Wave scenario~\cite{coherenceTimeRef}. 

	{\color{black}
		\subsection{Online Deployment of the Proposed Approach}
		\label{sec:OnlineDeploy}
		The adaptation of the DL network to changes in the propagation environment is critical because the offline training does not include all possible channel characteristics. To address this problem, we also adopt an online deployment strategy for F2 and F3 to perform both channel estimation and hybrid beamforming. 
		
		In our proposed online deployment, the channel is estimated via both DL and analytical methods. {\color{black}Thus, the proposed method can be regarded as a hybrid DL-analytical approach. However, the analytical method is infrequently used only when the DL network requires to be updated if the network's performance is worse than a predefined threshold.} As a result, the proposed approach trade-offs between the high computational complexity of the analytical approach by infrequent updates and the poor performance of the DL model due to the changes in the channel statistics. 
		Define the received signal at time $t$ for $m\in \mathcal{M}$ as $\mathbf{Y}_t[m]$ (see (\ref{receivedSignalPilotMod}) and $7$-th step of Algorithm~\ref{alg:algorithmTraining}). Then, the estimates of the channel matrix $\mathbf{H}_t[m]$ are obtained through DL as $\hat{\mathbf{H}}_t^\mathrm{DL}[m]$ and some analytical approach, such as the angle domain channel estimation (ADCE) algorithm~\cite{mimoAngleDomainFaiFai}, as $\hat{\mathbf{H}}_t^\mathrm{AD}[m]$.  
		Define the error metric at time $t$ as 
		\begin{align}
		\label{etaMetric3}
		\eta^{(t)} = \frac{1}{MN_\mathrm{R}N_\mathrm{T}}\sum_{m=1}^{M}|| \hat{\mathbf{H}}_t^\mathrm{DL}[m] - \hat{\mathbf{H}}^\mathrm{temp}[m]   ||_\mathcal{F}, 
		\end{align}
		where $\hat{\mathbf{H}}^\mathrm{temp} $ is the most recent estimated channel matrix via ADCE. Note that $\hat{\mathbf{H}}^\mathrm{temp} $ is not calculated at each iteration, instead it is only obtained after the employment of ADCE, hence, the term \emph{most recent} means the last time that the model update has been performed.  {\color{black}The channel estimation via ADCE is  performed infrequently only at the beginning of the online deployment and when the network is updated.} The update is performed by using the collected input data at time $t$ and the labels are obtained via ADCE. Consequently, 
		analytical solutions are used only when an update is required, thereby reducing the computational complexity, as illustrated in Table~\ref{tableOL}. The online training is also fast because a small dataset is used to update the network parameters for a few iterations. {\color{black}During the online deployment, the model update is performed when
			\begin{align}
			\label{etaMetric}
			\eta^{(t)} \geq \zeta,    
			\end{align}
			holds for some threshold parameter $\zeta$ which determines the refresh frequency and depends on the coherency of the channel; $\eta^{(t)} < \zeta$ implies satisfactory performance of DL.  The condition (\ref{etaMetric}) is checked by computing $\eta^{(t)}$ as in (\ref{etaMetric3}) for every channel estimate of DL method and $\hat{\mathbf{H}}^\mathrm{temp} $, which is conducted only when the network is updated. This is achieved by storing  $\hat{\mathbf{H}}^{\mathrm{temp}}[m]$ in the memory and use it to compute $\eta^{(t)}$.} 
		
		Denote the network parameters of MC-CENet at time $t$ as $\boldsymbol{\Pi}_t^{\mathrm{MC-CENet}}$; we use the same definition for SC-CENet$[m]$ and HBNet. {\color{black}In order to keep the learning capability of the trained network,} only the higher layers (i.e., the fully connected layers) of $\boldsymbol{\Pi}_t^{\mathrm{MC-CENet}}$, which have greater dependence on the environment than the lower layers (i.e., convolutional layers) are updated~\cite{elbir2020TL}. {\color{black}In practice, this is achieved by fixing the learning rate to zero for the lower layers.} {\color{black}This is because the lower layers, i.e., the convolutional layers, are generally domain-invariant. But it is not so with the higher layers which require re-training when new data and labels are added to the problem~\cite{elbir2020TL}.} This slight update of the network parameters does not cause significant performance loss in the input-output mapping that is already learned. In fact, it allows us to learn the new features in the environment. Algorithm~\ref{alg:OnlineDeploy} summarizes 	online deployment steps. The inputs are the received pilot signals and the offline trained networks. 		In online deployment, for labeling we use PE-HB (phase-extraction-based hybrid beamforming) algorithm~\cite{sohrabiOFDM} instead of MO due to its low complexity. {\color{black}Note that the analytical methods ADCE and PE-HB are used only when the network requires an update, thus, lowering the computational complexity of the approach.}
		
		\begin{algorithm}
			{\color{black}
				\begin{algorithmic}[1] \footnotesize
					\caption{Online deployment for MC-CENet and HBNet. }
					\Statex {\textbf{Input:} $\mathbf{Y}_0[m]$, $\boldsymbol{\Pi}_0^{\mathrm{MC-CENet}}$, $\boldsymbol{\Pi}_0^{\mathrm{HBNet}}$, $\zeta$.}
					\Statex	{\textbf{Output:} $\hat{\mathbf{H}}_t^\mathrm{DL}[m]$, $\hat{\mathbf{F}}_{\mathrm{RF}_t}$, $\hat{\mathbf{F}}_{\mathrm{BB}_t}[m]$, $\hat{\mathbf{W}}_{\mathrm{RF}_t}$, $\hat{\mathbf{W}}_{\mathrm{BB}_t}[m]$.}
					\label{alg:OnlineDeploy}
					\Statex \textbf{Initialization} 
					\State Start with $t=0$. 
					\State Estimate $\mathbf{H}_t[m]$ via  $\boldsymbol{\Pi}_t^{\mathrm{MC-CENet}}$ and ADCE as $\hat{\mathbf{H}}_t^{\mathrm{DL}}[m]$ and  $\hat{\mathbf{H}}_t^{\mathrm{AD}}[m]$.
					\State Set $\hat{\mathbf{H}}^\mathrm{temp}[m] = \hat{\mathbf{H}}_t^{\mathrm{AD}}[m]$ {\color{black}and store it in the memory}.
					\State Compute the error metric $\eta^{(t)}$ in (\ref{etaMetric3}).
					\Statex \textbf{Repeat} 
					\State \indent \textbf{if} $\eta^{(t)} < \zeta$,  \textbf{\color{black}[No Model Update]} 
					\State \indent  Estimate $\mathbf{H}_t[m]$ via $\boldsymbol{\Pi}_t^{\mathrm{MC-CENet}}$  as $\hat{\mathbf{H}}_t^{\mathrm{DL}}[m]$.
					
					\State \indent Compute $\eta^{(t)}$ in (\ref{etaMetric3}).
					\State \indent Obtain $\hat{\mathbf{F}}_{\mathrm{RF}_t}$, $\hat{\mathbf{F}}_{\mathrm{BB}_t}[m]$, $\hat{\mathbf{W}}_{\mathrm{RF}_t}$, $\hat{\mathbf{W}}_{\mathrm{BB}_t}[m]$ via $\boldsymbol{\Pi}_t^{\mathrm{HBNet}}$.
					\State \indent $\boldsymbol{\Pi}_{t+1}^{\mathrm{MC-CENet}}= \boldsymbol{\Pi}_{t}^{\mathrm{MC-CENet}}$, $\boldsymbol{\Pi}_{t+1}^{\mathrm{HBNet}}= \boldsymbol{\Pi}_{t}^{\mathrm{HBNet}}$.
					\State \indent $t \leftarrow t + 1$.
					\State \indent \textbf{else}, \textbf{\color{black}[Model Update]}
					\State \indent For $m\in\mathcal{M}$, estimate $\mathbf{H}_t[m]$ via  ADCE as  $\hat{\mathbf{H}}_t^{\mathrm{AD}}[m]$.
					\State \indent Set $\hat{\mathbf{H}}^\mathrm{temp}[m] = \hat{\mathbf{H}}_t^{\mathrm{AD}}[m]$ {\color{black}and store $\hat{\mathbf{H}}^\mathrm{temp}[m]$ in the memory}.
					\State \indent Use $\mathbf{Y}_t[m]$ and $\hat{\mathbf{H}}_t^{\mathrm{AD}}[m]$ to generate online dataset  $\overline{\mathcal{D}}_\mathrm{MC-CENet}$
					as in $12$-th step of  Algorithm~\ref{alg:algorithmTraining}.
					\State \indent Freeze the convolutional layers of $\boldsymbol{\Pi}_t^\mathrm{MC-CENet}$ and $\boldsymbol{\Pi}_{t}^{\mathrm{HBNet}}$.
					\State \indent Update $\boldsymbol{\Pi}_t^\mathrm{MC-CENet}$ with $\overline{\mathcal{D}}_\mathrm{MC-CENet}$ and obtain  $\boldsymbol{\Pi}_{t+1}^\mathrm{MC-CENet}$.
					\State \indent Estimate  $\mathbf{H}_t[m]$ via $\boldsymbol{\Pi}_{t+1}^\mathrm{MC-CENet}$ and refine $\hat{\mathbf{H}}_t^{\mathrm{DL}}[m]$.
					\State \indent {\color{black}Compute $\eta^{(t)}$ in (\ref{etaMetric3})}.
					\State \indent Using $\hat{\mathbf{H}}_t^{\mathrm{DL}}[m]$, estimate $\hat{\mathbf{F}}_{\mathrm{RF}_t}$, $\hat{\mathbf{F}}_{\mathrm{BB}_t}[m]$, $\hat{\mathbf{W}}_{\mathrm{RF}_t}$,  $\hat{\mathbf{W}}_{\mathrm{BB}_t}[m]$ via PE-HB algorithm.
					\State \indent Construct online dataset $\overline{\mathcal{D}}_\mathrm{HBNet}$ as in $13$-th  steps of Algorithm~\ref{alg:algorithmTraining}.
					\State \indent Freeze the convolutional layers of $\boldsymbol{\Pi}_t^\mathrm{HBNet}$.
					\State \indent Update $\boldsymbol{\Pi}_t^\mathrm{HBNet}$ with $\overline{\mathcal{D}}_\mathrm{HBNet}$ and obtain $\boldsymbol{\Pi}_{t+1}^\mathrm{HBNet}$.
					\State \indent Estimate  beamformers via $\boldsymbol{\Pi}_{t+1}^\mathrm{HBNet}$.
					\State \indent $t \leftarrow t + 1$.
					\State \indent \textbf{end}
					\Statex \textbf{Until} Algorithm shutdown
				\end{algorithmic}
			}
		\end{algorithm}
		
		\subsection{Computational Complexity and Power Consumption}
		\label{sec:CComp}
		The computational complexity of the proposed DL approach arises from both online prediction and offline training. While the online complexity is easier to compute, the same for offline training is still an open issue due to a more involved implementation of backpropagation process during training~\cite{complexityOfflineTraining}. Therefore, we consider the complexity of only online prediction stage here. 
		
		For a deep neural network with $L_\mathrm{C}$ convolutional layers and $L_\mathrm{F}$ fully connected layers, the total time complexity of convolutional layers is $\mathcal{O}( \sum_{l=1}^{L_\mathrm{C}}D_x^{(l)}D_y^{(l)} b_x^{(l)}b_y^{(l)}c_\mathrm{CL}^{(l-1)}c_\mathrm{CL}^{(l)}   )$, where $D_x^{(l)}$ and $D_y^{(l)}$ are the convolutional kernel dimensions; $ b_x^{(l)}$ and $b_y^{(l)}$ are the dimensions of the $l$-th convolutional layer output; and $c_\mathrm{CL}^{(l)}$ is the number of filters in the $l$-th layer. The total complexity of the fully connected layers is $\mathcal{O} (\sum_{l=1}^{L_\mathrm{F}}b_x^{(l)} b_y^{(l)} c_\mathrm{F}^{(l)}  ) $, where $c_\mathrm{FL}^{(l)}$ is the number of units in the fully connected layer. A fair comparison of computational complexity for a DL network with a conventional analytical method is difficult because DL enjoys advantage of GPU processing. In Section~\ref{sec:complexitySim}, we compare the computation times of both methods.
	}
	
	\begin{figure*}[t]
		\centering
		\subfloat[]{\includegraphics[draft=false,width=.35\textheight]{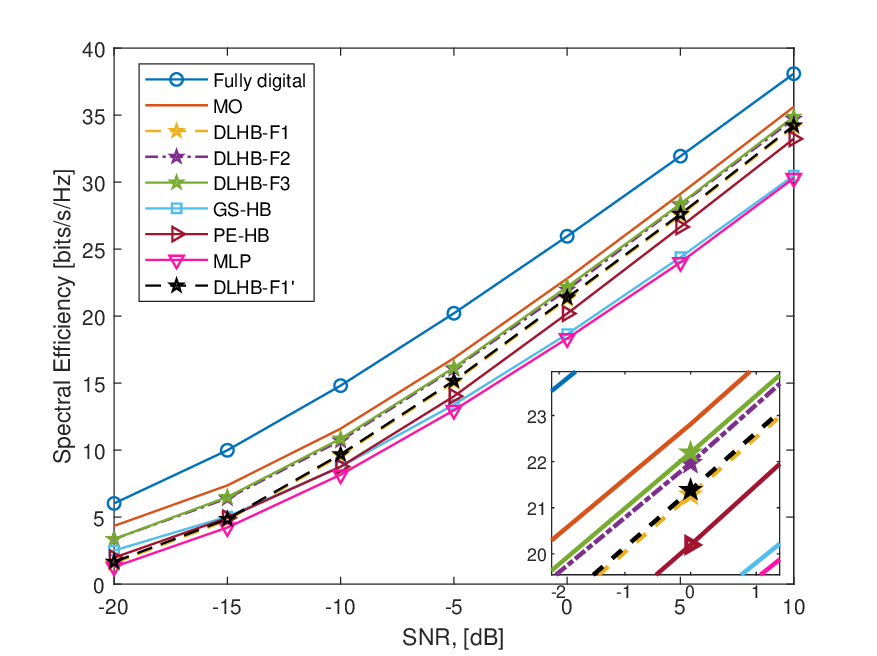} } 
		\subfloat[]{\includegraphics[draft=false,width=.35\textheight]{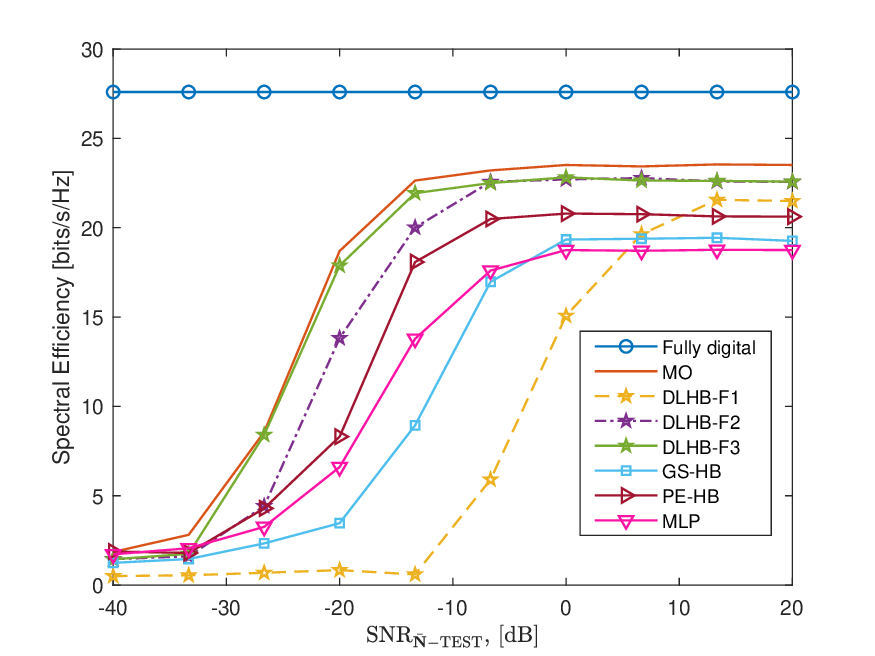} } 
		\caption{(a) Spectral efficiency versus SNR. $N_\mathrm{T} = 128$, $N_\mathrm{R}=16$, $M=16$ and  SNR$_{\overline{\mathbf{N}}-\mathrm{TEST}}=20$ dB. Inset shows the plot on a magnified scale. (b) Spectral efficiency versus SNR$_{\overline{\mathbf{N}}-\mathrm{TEST}}$ and SNR $=0$ dB.    }
		\label{fig_SNR_Rate}
	\end{figure*}
	
	{\color{black} Next, we examine power consumption of the proposed DL approach~\cite{energyEfficiency_Bjorn}. For a conventional mm-Wave system with transmit power $P_\mathrm{T}=1$ W, total power consumption is
		\begin{align}
		P_\mathrm{tot} = P_\mathrm{T} + N_\mathrm{RF}P_\mathrm{RF} + N_\mathrm{T}N_\mathrm{RF} P_\mathrm{PS} + P_\mathrm{BB},
		\end{align}
		where $ P_\mathrm{RF} $, $P_\mathrm{PS}$ and $P_\mathrm{BB}$ are power consumption of RF chain, phase shifter and baseband processing, respectively. Approximately, $ P_\mathrm{RF} =250$ mW, $P_\mathrm{PS}=40$ mW and $P_\mathrm{BB}=200$ mW. This yields $P_\mathrm{tot} = 22.68$ W for $N_\mathrm{T}=128$ transmit antennas and $N_\mathrm{RF}=4$ RF chains~\cite{mimoChannelModel2}. 
		While there is no commercial DL-based hardware for mm-Wave system configuration, there exist some processing units that run DL methods effectively. For example, Intel Movidius has power consumption of approximately $500$ mW. In addition, a DL-based prototype for mm-Wave communication is proposed in \cite{dl_prototype_5G}, wherein DL-based methods are shown to be promising for next generation wireless communication systems that exhibit power efficiency and low computation complexity.}

	\section{Numerical Simulations}
	\label{sec:Sim}
	We evaluated the performance of the proposed DL frameworks through several experiments. We compared our DL-based hybrid beamforming (hereafter, DLHB) with the state-of-the-art hybrid precoding algorithms such as  Gram-Schmidt-orthogonalization-based method (GS-HB) \cite{alkhateeb2016frequencySelective}, PE-HB \cite{sohrabiOFDM}, and another recent DL-based multilayer perceptron (MLP) method \cite{mimoDeepPrecoderDesign}. In addition, we compare the channel estimation performance with ADCE, least squares (LS) and minimum MSE (MMSE). As a benchmark, we implemented a fully digital beamformer obtained from the SVD of the channel matrix. We also present the performance of the MO algorithm \cite{hybridBFAltMin} used for the labels of the hybrid beamforming networks. The MO algorithm constitutes a performance yardstick for DLHB, in the sense that the latter cannot perform better than the MO algorithm because the hybrid beamformers used as labels are obtained from MO itself. Finally, we implemented spatial frequency CNN (SF-CNN) architecture \cite{deepCNN_ChannelEstimation} that has been proposed recently for wideband mm-Wave channel estimation. We compare the performance of our DL-based channel estimation with SF-CNN using the same parameters.

	
	\begin{figure*}[h]
		\centering
		\subfloat[]{\includegraphics[draft=false,width=.265\textheight]{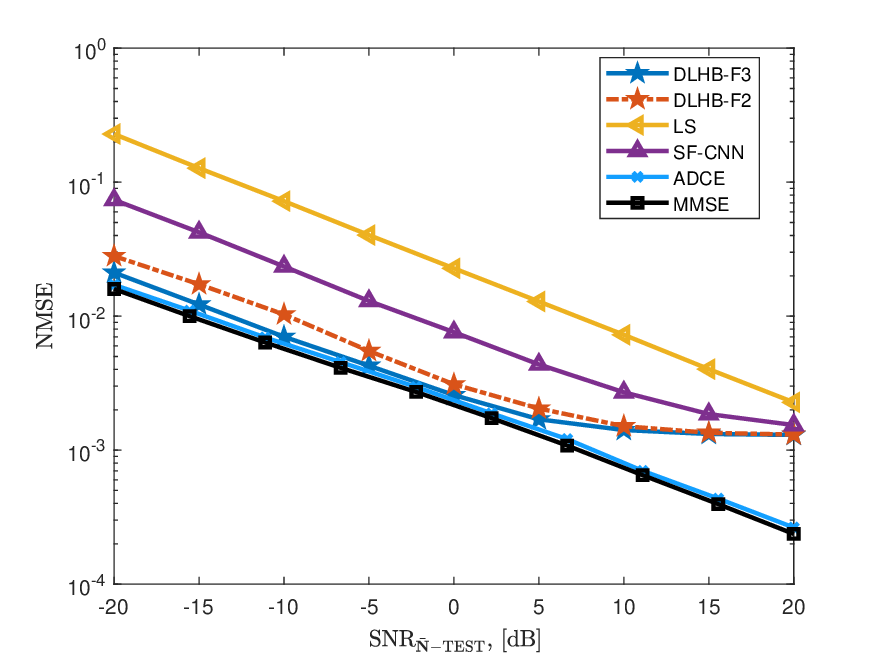} } 
		\subfloat[]{\includegraphics[draft=false,width=.265\textheight]{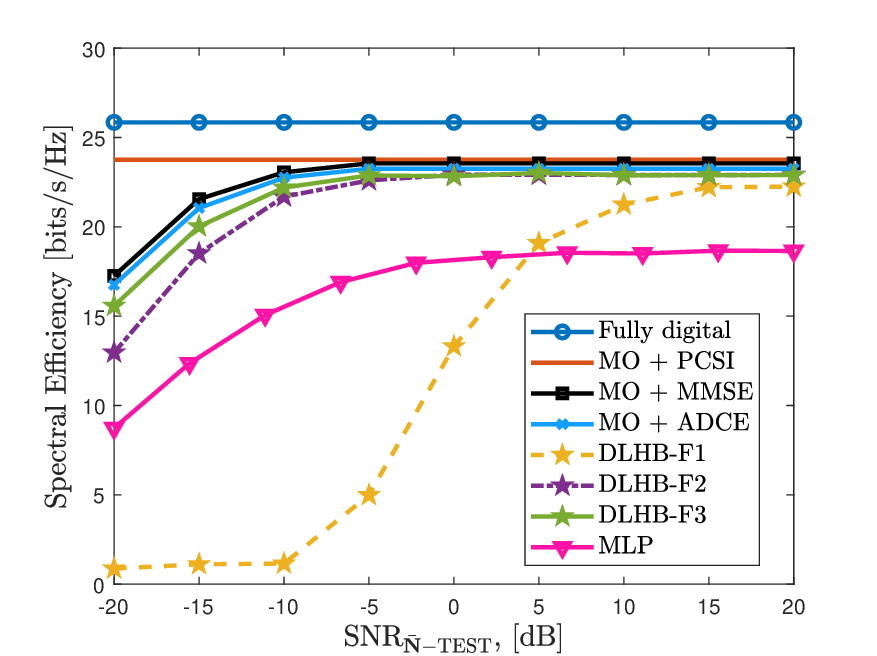} }
		\\ 
		\subfloat[]{\includegraphics[draft=false,width=.265\textheight]{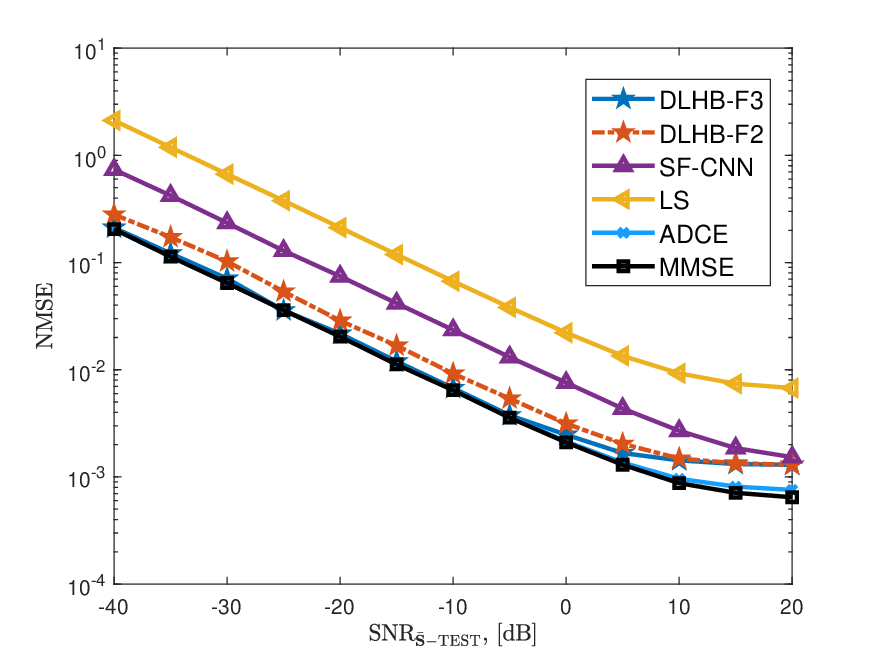} } 
		\subfloat[]{\includegraphics[draft=false,width=.265\textheight]{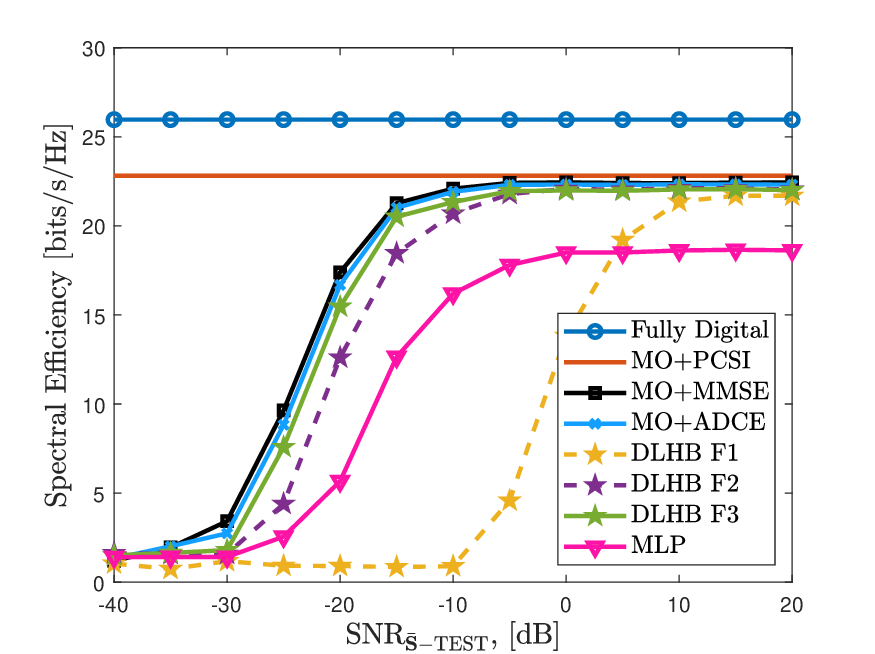} } 
		\caption{Channel estimation performance of DL approaches with respect to  SNR$_{\overline{\mathbf{N}}-\mathrm{TEST}}$ in terms of (a) NMSE and (b) spectral efficiency for SNR $=0$ dB.    Channel estimation performance with respect to SNR$_{\overline{\mathbf{S}}-\mathrm{TEST}}$ in terms of (c) NMSE and (d) spectral efficiency for  SNR$_{\overline{\mathbf{N}}-\mathrm{TEST}}= 10$ dB.}
		\label{fig_CE_SNRonReceivedSignal}
	\end{figure*}

	\begin{figure}[h]
		\centering
		{\includegraphics[draft=false,width=.5\textwidth]{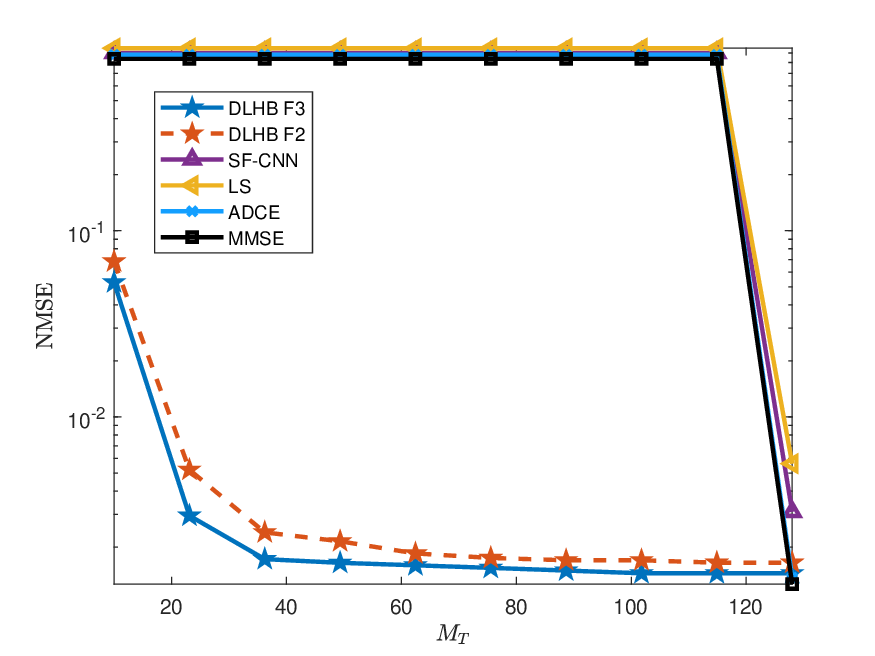} } 
		
		\caption{\color{black}Channel estimation performance with respect to number of pilot signals $M_\mathrm{T}$ for  SNR$_{\overline{\mathbf{N}}-\mathrm{TEST}}= 10$ dB.}
		\label{fig_CE_Mt}
	\end{figure}

	The proposed deep networks are realized and trained in MATLAB on a PC with a single GPU and a 768-core processor. We used the stochastic gradient descent algorithm with momentum 0.9  and  updated the network parameters with learning rate $0.0005$ and mini-batch size of $128$ samples. Then, we reduced the learning rate by the factor of $0.9$ after each 30 epochs. {\color{black} The training process terminates when the validation accuracy does not improve in three consecutive epochs.} Algorithm~\ref{alg:algorithmTraining} summarizes steps for training data generation. 
	We followed the training procedure outlined in the Section~\ref{sec:HD_Design} with $N_\mathrm{T}=128$ elements, $N_\mathrm{R}=16$ antennas, and $N_\mathrm{RF} = N_\mathrm{S} = 4$ RF chains.	Throughout the experiments, unless stated otherwise, we use $M=16$ subcarriers at $f_c = 60$ GHz with $2$ GHz bandwidth, and $L=10$ clusters with $N_\mathrm{sc}=5$ scatterers for $M_\mathrm{T}=64$ and $M_\mathrm{R}=16$ and for all transmit and receive angles that are uniform randomly selected from the interval $[-\pi,\pi]$. 
	{\color{black} Then the frequency selective channel is generated by following the channel model in (\ref{eq:delaydChannelModel}) and (\ref{Hm_OFDM})~\cite{alkhateeb2016frequencySelective}. }
	In the prediction stage, the preamble data are different from the training stage. Instead, we construct $\mathbf{Y}[m]$ from (\ref{receivedSignalPilot}) with a completely different realization of noise $\overline{\mathbf{N}}$ corresponding to SNR$_{\overline{\mathbf{N}}-\mathrm{TEST}}$. 

	\subsection{Spectral efficiency evaluation}
	Fig.~\ref{fig_SNR_Rate} shows the spectral efficiency of various algorithms for varying test SNR, given SNR$_{\overline{\mathbf{N}}}=20$ dB. {\color{black}Note that only DLHB is fed with the received pilot data (i.e., $\mathbf{Y}[m]$) whereas the other algorithms require perfect CSI to yield hybrid beamformers. Nevertheless,} the DLHB techniques outperform GS-HB \cite{alkhateeb2016frequencySelective} and PE-HB \cite{sohrabiOFDM}.  Further, GS-HB algorithm requires the set of array responses of received paths which is difficult to achieve in practice. The MO algorithm is used to obtain the labels of the deep networks for hybrid beamforming, hence the performances of the DL approaches are upper-bounded by the MO algorithm. {\color{black}The perfect channel information is required for even the benchmark MO algorithm \cite{hybridBFAltMin}}. The gap between the MO algorithm and the DL frameworks is explained by the corruptions in the DL input which causes deviations from the label data (obtained via MO) at the output regression layer. Note that our DLHB methods improve upon other DL-based techniques such as MLP \cite{mimoDeepPrecoderDesign}, which lacks a feature extraction stage provided by convolutional layers in our networks. {\color{black}Among the DL frameworks, F2 and F3 exhibit superior performance than F1 because the channel estimated by MC-CENet and SC-CENet has higher accuracy. On the contrary, F1 uses $\mathbf{Y}[m]$ directly as input and is, therefore, unable to achieve similar improvement because of the absence of channel estimation stage.} \textcolor{black}{We further present the performance of F1 with deeper learning model, denoted as F1$^{'}$, which includes two additional convolutional layers of the same size as in F1. We observe that deeper learning architecture of F1$^{'}$ provides an incremental performance improvement in spectral efficiency.} While F2 and F3 have similar hybrid beamforming performance, F3 has computationally more complex because of presence of $M$ CNNs in the channel estimation stage.

	{\color{black}In Fig.~\ref{fig_SNR_Rate}(a), except DLHB, all competing methods use perfect CSI except for DLHB. For a fair comparison, we feed all methods the same channel matrix that is estimated from MC-CENet when SNR $=0$ dB.} Fig.~\ref{fig_SNR_Rate}(b) shows the spectral efficiency so obtained with respect to SNR$_{\overline{\mathbf{N}}-\mathrm{TEST}}$, {\color{black}which determines the noise added to the received pilot data as in Section~\ref{sec:NetTraining}}. For SNR$_{\overline{\mathbf{N}}-\mathrm{TEST}}\geq 0$ dB, we note that the non-DL methods perform rather imperfectly but their performance is at least similar with the true channel matrix case shown in Fig~\ref{fig_SNR_Rate}(a). The DL-based techniques exceed in comparison and exhibit higher tolerance against the corrupted channel data corresponding to SNR$_{\overline{\mathbf{N}}-\mathrm{TEST}}$. The F2 and F3 quickly reach the maximum efficiency when SNR$_{\overline{\mathbf{N}}-\mathrm{TEST}}$ is increased to $-15$ dB. Again, the F1 fares poorly  because it is directly fed by the received data and lacks the channel estimation network.


	\begin{figure}[h]
		\centering
		\subfloat[]{\includegraphics[draft=false,width=.255\textheight]{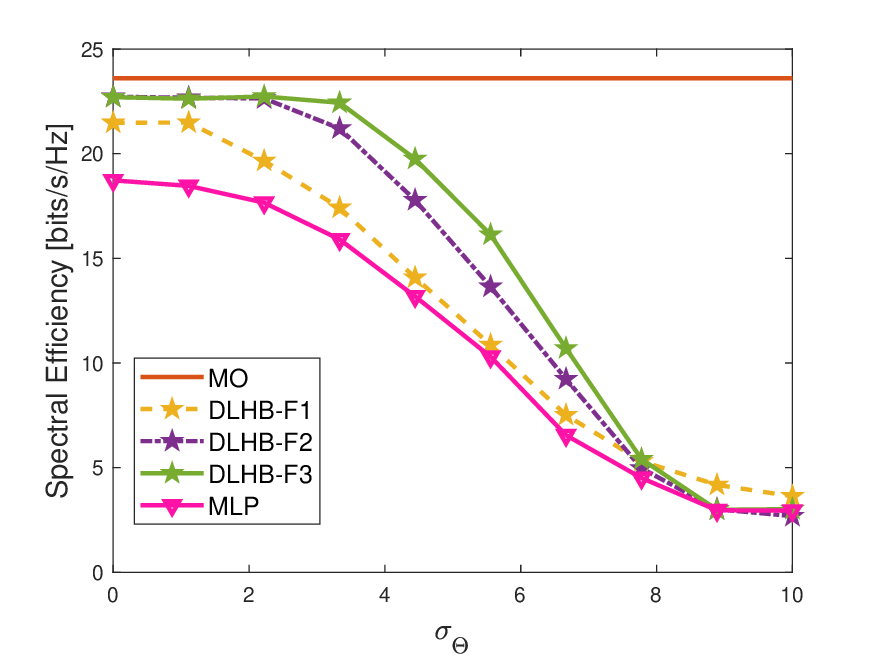} }\\
		
		\subfloat[]{\includegraphics[draft=false,width=.255\textheight]{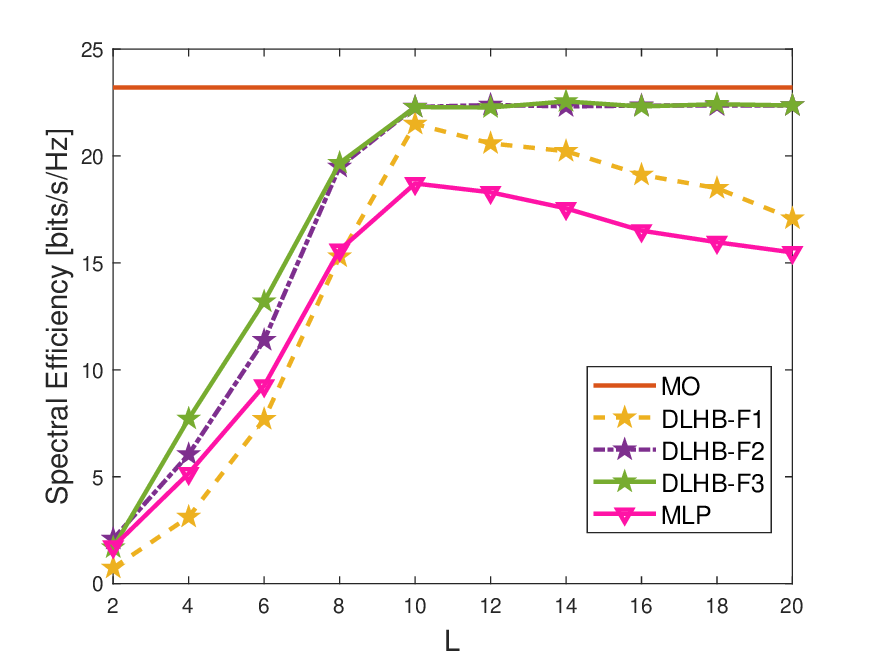} } 
		
		\caption{(a) Performance of DL approaches with respect to the standard deviation of angle mismatch when  $L= 10$ for SNR $=0$ dB and SNR$_{\overline{\mathbf{N}}-\mathrm{TEST}}= 10$ dB. (b) Performance with respect to number of cluster mismatch when the networks are trained for $L= 10$.}
		\label{fig_AngleMismatch}
	\end{figure}

	\subsection{Error in channel estimation}
	\label{subsec:ch_est}

	
	Fig.~\ref{fig_CE_SNRonReceivedSignal} shows the normalized MSE (NMSE) (Fig.~\ref{fig_CE_SNRonReceivedSignal}(a)) in the channel estimates and the spectral efficiency (Fig.~\ref{fig_CE_SNRonReceivedSignal}(b)) of the DL approaches with respect to  SNR$_{\overline{\mathbf{N}}-\mathrm{TEST}}$ when SNR $=0$ dB. Here, the NMSE is $	\textrm{NMSE} = \frac{1}{M J_T }  \sum_{m=1}^{M}\sum_{i=1}^{J_T}  \frac{|| \mathbf{H}[m] - \hat{\mathbf{H}}_i[m] ||_\mathcal{F}}{|| \mathbf{H}[m]  ||_\mathcal{F}   } ,$
	where $J_T$ is the number trials.  
	We observe that all of the  DL frameworks provide improvement as SNR$_{\overline{\mathbf{N}}-\mathrm{TEST}}$ increases but F3, in particular, surpasses all other methods. We remark that DLHB approaches outperform the recently proposed SF-CNN because the latter lacks fully connected layers and relies only on several convolutional layers (see Table 1 in \cite{deepCNN_ChannelEstimation}). While convolutional layers are good at extracting the additional features inherent in the input, the fully connected layers are more efficient in non-linearly mapping the input to the labeled data. Further, SF-CNN~\cite{deepCNN_ChannelEstimation} draws on a single SNR$_{\overline{\mathbf{N}}}$ in the training and works well only when SNR$_{\overline{\mathbf{N}}}=$ SNR$_{\overline{\mathbf{N}}-\mathrm{TEST}}$. This is impractical because it requires re-training whenever there is a change in SNR$_{\overline{\mathbf{N}}-\mathrm{TEST}}$. On the other hand, no such requirement is imposed on our DLHB method because we use multiple SNR$_{\overline{\mathbf{N}}}$s during the training stage. {\color{black}In addition, The proposed DL networks accepts all the subcarrier data jointly, thus, providing better feature extraction whereas SF-CNN sequentially processes the subcarrier data by accepting only two channel matrices at a time.} Again, F3 leverages multiple CNNs to outclass F2. While both have largely similar results as in Fig.~\ref{fig_SNR_Rate}(a), we observe from Fig.~\ref{fig_CE_SNRonReceivedSignal}(b) that F3 attains higher spectral efficiency even at SNR$_{\overline{\mathbf{N}}-\mathrm{TEST}}$ as low as -5 dB when compared with F1, F2, and MLP. We conclude that, effectively, the channel estimation improvement in F3 also leads to capacity enhancement at very low SNR$_{\overline{\mathbf{N}}-\mathrm{TEST}}$. 
	
	Next, Fig.~\ref{fig_CE_SNRonReceivedSignal}(b) illustrates that F1 performs well only when SNR$_{\overline{\mathbf{N}}-\mathrm{TEST}}$ exceeds 15 dB. In summary, F2 yields the highest spectral efficiency with reasonable network complexity. We observe in Fig.~\ref{fig_CE_SNRonReceivedSignal}(a) that the performance of DL-based algorithms maxes out  after SNR$_{\overline{\mathbf{N}}-\mathrm{TEST}}$ reaches $5$ dB. This is because, being biased estimators, deep networks do not provide unlimited accuracy. This problem can be mitigated by increasing the number of units in various network layers. Unfortunately, it may lead to the network memorizing the training data and perform poorly when the test data are different than the ones in training. To balance this trade-off, we used noisy data-sets during training so that the network attains reasonable tolerance to corrupted/imperfect inputs. Although the spectral efficiency of DLHB frameworks remains largely unchanged at high SNR$_{\overline{\mathbf{N}}-\mathrm{TEST}}$, it is an improvement over MLP as can be ascertained from both Fig.~\ref{fig_SNR_Rate} and Fig.~\ref{fig_CE_SNRonReceivedSignal}(b).

	{\color{black} To examine the performance for different number of pilot signals, we present the channel estimation performance with respect to  $M_\mathrm{T}$ in Fig.~\ref{fig_CE_Mt}, $M_\mathrm{T}=N_\mathrm{T}$ is selected in Fig.~\ref{fig_CE_SNRonReceivedSignal}. We can see that the model-based methods, such as LS, ADCE and MMSE, as well as SF-CNN suffer from the loss of pilot signals. However, the proposed DL approach provides significantly better NMSE even if $M_\mathrm{T}<N_\mathrm{T}$ and reliable NMSE can be achieved by F2 and F3 for as low pilots as $M_\mathrm{T}=40$. The outperformance of DLHB can be attributed to the use of raw data instead of preprocessed data as in SF-CNN, whose performance significantly degrades due to matrix inversion. }

	
	\subsection{Effect of noise contamination} 
	We examined the performance of the DL approaches for the corrupted pilot data when SNR $=0$ dB and SNR$_{\overline{\mathbf{N}}-\mathrm{TEST}}= 10$ dB. In this experiment, we added noise determined by SNR$_{\overline{\mathbf{S}}-\mathrm{TEST}}$ to the pilot signal matrix $\overline{\mathbf{S}}$ in (\ref{receivedSignalPilot}). All networks are trained by selecting $\overline{\mathbf{S}} = \sqrt{P_T} \mathbf{I}_{M_\mathrm{T}}$. 
	Fig.~\ref{fig_CE_SNRonReceivedSignal}(c) shows that F3 has lower NMSE than both F2 and SF-CNN. Here, the performance of the algorithms maxes out after SNR$_{\overline{\mathbf{S}}-\mathrm{TEST}}$ is increased to $15$ dB; the channel estimation improvement is very incremental for all deep networks except LS, where the preamble noise is determined by SNR$_{\overline{\mathbf{N}}-\mathrm{TEST}}$. The degradation in accuracy of DL methods can be similarly explained as in Section~\ref{subsec:ch_est}. Nevertheless, the hybrid beamforming performance of F2 and F3 is better than MLP even though the channel estimation improvement is modest. Moreover, the performance of F2 and F3 quickly reaches to their best after SNR$_{\overline{\mathbf{S}}-\mathrm{TEST}} = -15$ dB (Fig.~\ref{fig_CE_SNRonReceivedSignal}(d)).
	
	\begin{figure*}[t]
		\centering
		\subfloat[]	{\includegraphics[draft=false,width=.5\linewidth]{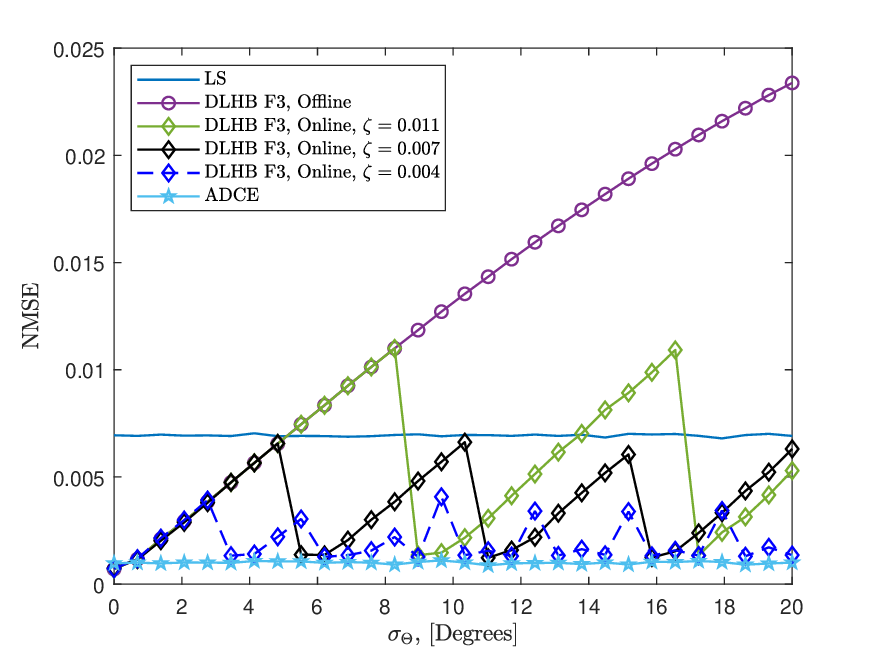} }
		\subfloat[]	{\includegraphics[draft=false,width=.5\linewidth]{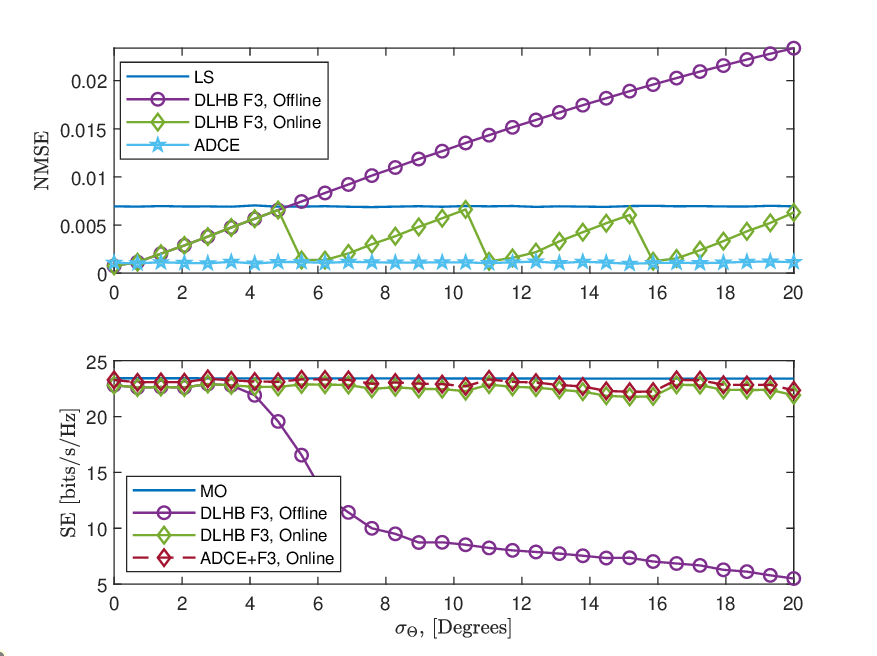} }
		\caption{\color{black} Online Deployment of the proposed DL approaches. (a) Channel estimation NMSE for $\zeta \in \{0.004, 0.007, 0.011\}$. (b) Channel estimation NMSE and  spectral efficiency and with respect to angular mismatch for $\sigma_\Theta\in [0^\circ,20^\circ]$ when SNR $=0$ dB.}

		\label{fig_Online0_20}
	\end{figure*}

	\begin{figure}[t]
		\centering
		{\includegraphics[draft=false,width=\columnwidth]{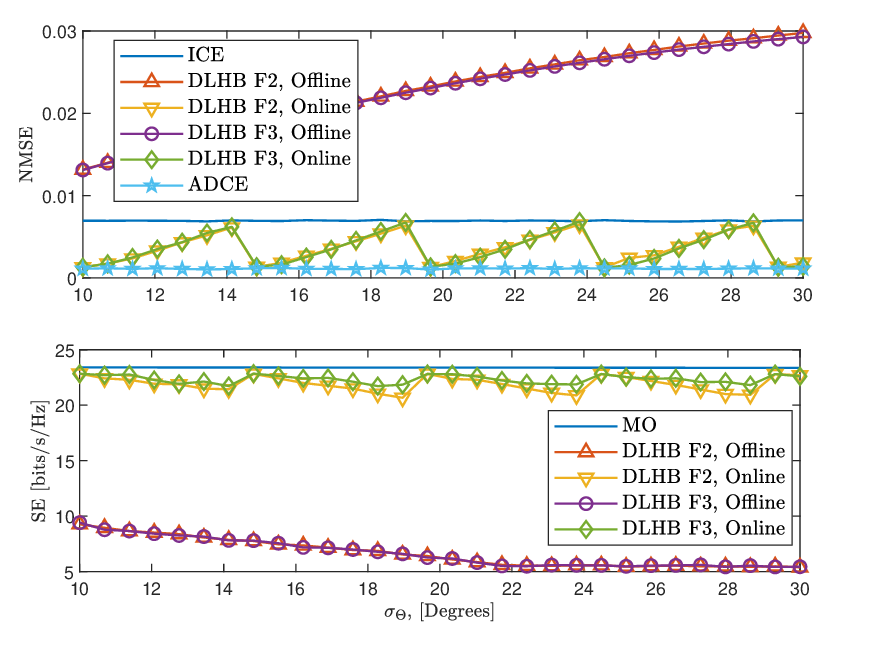} }
		\caption{\color{black} Online channel estimation performance of the proposed DL approach for $\sigma_\Theta\in [10^\circ,30^\circ]$. (Top) channel estimation NMSE (Middle) spectral efficiency and (Bottom) error cost and threshold with respect to angular mismatch for  SNR $=0$ dB.}
		\label{fig_Online10_30}
	\end{figure}

	\subsection{Effect of angle and cluster mismatch}
	\label{sec:angleMismatch}
	We imposed further challenges on our techniques by introducing an angle mismatch from the receiver AOA/AOD angles (also used as training data). In the prediction stage, we generated a different channel matrix by inserting angular mismatch in each of the path angles. Fig.~\ref{fig_AngleMismatch}(a) illustrates the spectral efficiency achieved with respect to the standard deviation of the mismatch angle, $\sigma_\Theta$ in degrees. Hence, for the AOA/AOD angles ${\theta}_l,{\phi}_l$ from the $l$-th cluster, the mismatched angles are given by $\widetilde{\theta}_l \sim \mathcal{N}(\theta_l, \sigma_\Theta^2)$ and $\widetilde{\phi}_l \sim \mathcal{N}(\phi_l, \sigma_\Theta^2)$, respectively. 
	For $L=10$ clusters, DLHB methods are able to tolerate at most $5^{\circ}$ of angular mismatch which other learning-based methods such as MLP are unable to. As this mismatch increases, it leads to significant deviations in the channel matrix data (arising from the multiplication of deviated steering vectors in (\ref{eq:delaydChannelModel}).
	
	We also evaluated effect of a mismatch in the number of clusters $L$ between training and prediction data. We trained the networks for $L=10$with different channel realizations. During testing, we generated a new channel matrix for different number of clusters. Figures~\ref{fig_AngleMismatch}(b) illustrates the spectral efficiency for $L=10$. The F2 and F3 reach to their maximum performance when $L$ reaches to the value used in the training. The performance of F1 and MLP gets worse as $L$ increases.	Note that in the prediction stage, the first 10 cluster angles, as in Fig.~\ref{fig_AngleMismatch}(b), are same as used for training; remaining 10 cluster angles are selected uniformly at random as mentioned earlier. As $L$ increases, the input data becomes ``more familiar'' to the deep network. The spectral efficiency does not degrade after addition of randomly generated cluster paths because DLHB designs the hybrid beamformers according to the received paths that are already present in the training data. As a result, deep networks provide robust performance even with additional received paths and channel matrix different from the training stage. However, the loss of cluster paths in the training data  would deteriorate the performance because the input data becomes ``unfamiliar'' to the deep network and hybrid beamformer designs suffer as a result. Hence, it is recommended to select $L$ small, approximately $L\in [3,7]$ so that the DL network can handle the scenario when $L$ is higher.
	
	\begin{table*}[t]
		\caption{ Training Times for Networks (in Minutes) }
		\label{tableComp_Networks}
		\centering
		\begin{tabular}{|c|c|c|c|c|c|c|c|c|}
			\hline
			\hspace{-3pt}		MC-HBNet				\hspace{-3pt} &\hspace{-3pt} MC-CENet													\hspace{-3pt}& \hspace{-3pt}SC-CENet$[m]$\hspace{-3pt}&\hspace{-3pt} HBNet \hspace{-3pt}& \hspace{-3pt}SF-CNN\hspace{-3pt} &\hspace{-3pt} MLP\hspace{-3pt}&DLHB-F1			 & DLHB-F2												& DLHB-F3\\
			\hline
			45.6 &	95.3 & 	76.6	& 43.8 & 85.1& 41.4& 45.6&	138.5 & 	1270.3	\\
			\hline 
		\end{tabular}
	\end{table*}
	%
	
	\begin{table*}[t]
		\caption{Run Times for Algorithms (in Seconds) }
		\label{tableComp_Networks2}
		\centering
		{\color{black}
			\begin{tabular}{|c|c|c|c|c|c|c|c|}
				\hline
				\hspace{-3pt}MC-HBNet	\hspace{-3pt}			 & \hspace{-3pt}MC-CENet\hspace{-3pt}			& \hspace{-3pt}SC-CENet$[m]$\hspace{-3pt}&\hspace{-3pt} HBNet\hspace{-3pt} & \hspace{-3pt}SF-CNN\hspace{-3pt}&\hspace{-3pt}MLP\hspace{-3pt}&\hspace{-3pt}ADCE\hspace{-3pt} &\hspace{-3pt}MMSE\hspace{-3pt}\\
				\hline
				0.0053 &	0.0059 & 	0.0057	& 0.0059 & 0.0057 & 0.0056&0.010 &0.007\\
				\hline
				DLHB-F1			 & DLHB-F2												& DLHB-F3 &PE-HB& GS-HB &MO &\hspace{-3pt}ADCE+HBNet\hspace{-3pt} &ADCE+MO\\
				\hline
				0.0053&	0.0116 & 	0.0778& 0.0152 & 0.0132  &	3.204&0.0156 &3.214\\
				\hline 
		\end{tabular}}
	\end{table*}
	
	\begin{table*}[h!]
		\caption{Training Complexity over $1000$ Coherence Times (in Seconds) }
		\label{tableOL}
		\centering 
		\begin{tabular}{|p{0.16\textwidth} |p{0.16\textwidth}|  p{0.16\textwidth} |p{0.35\textwidth}| }
			\hline
			&Analytical Approach			 & DL, Offline	& DL, Online\\
			\hline
			Complexity order &$1000T_\mathrm{AA}$  &	$1000 T_\mathrm{DL}$ & $5T_\mathrm{AA} + 5T_\mathrm{OT}$ + $995T_\mathrm{DL}$ 	\\
			\hline
			Time complexity &  $1000\times 25\times 10^{-3}= 25$  &  $1000\times 0.0115=11.5$  & $5\times 25\times 10^{-3}+ 5\times 0.5 +  995\times 0.115
			= 14.0675$         \\
			\hline 
		\end{tabular}
	\end{table*}
	
	{\color{black}
		
		\subsection{Online Deployment}
		Fig.~\ref{fig_Online0_20} shows the online prediction performance of F3 when there is an angle mismatch between the AOD/AOA angles of training and test data. After experimental trial, we select the size of online dataset as $G=200$. We use ADCE~\cite{mimoAngleDomainFaiFai} and PE-HB~\cite{sohrabiOFDM} algorithms to obtain the online labels (Algorithm~\ref{alg:OnlineDeploy}). {\color{black}Fig.~\ref{fig_Online0_20}(a), we model the receiver motion such that $\sigma_\Theta\in [0^\circ, 20^\circ]$ for $\zeta \in \{0.004,0.007,0.011\}$ to demonstrate the performance with respect to the threshold parameter $\zeta$. As it is seen $\zeta$ determines the frequency of the model update; if $\zeta$ is small, then the neural network is updated more frequently and vice versa. We can conclude that $\zeta = 0.007$ is a good choice such that DL model does not perform worse than LS, but exhibits a much lower complexity.}		The performance of the offline network worsens as $\sigma_\Theta$ increases because it does not recognize the mismatched inputs. \textcolor{black}{When the online mode is triggered from a large angle mismatch (about $10^\circ$), Fig.~\ref{fig_Online10_30} shows the network updates itself iat the beginning of the deployment. Note that its performance is similar to Fig.~\ref{fig_Online0_20}.}
		The network requires to be re-trained for approximately every $5^\circ$ mismatch, similar to the observations made in Section~\ref{sec:angleMismatch}.
		
		The online deployment requires approximately $10$ and $15$ ms for channel estimation (ADCE) and beamformer design (PE-HB), respectively.  The online network update (training) takes only $500$ ms whereas the offline training overhead is about two hours (see Tables~\ref{tableComp_Networks}) mainly due to small online dataset ($\overline{\mathcal{D}}_\mathrm{MC-CENet}$ and $\overline{\mathcal{D}}_\mathrm{HBNet}$). 

	}
	
	%
	
	\subsection{Computational complexity}	
	\label{sec:complexitySim}

	\begin{figure}[ht]
		\centering
		{\includegraphics[draft=false,width=.35\textheight]{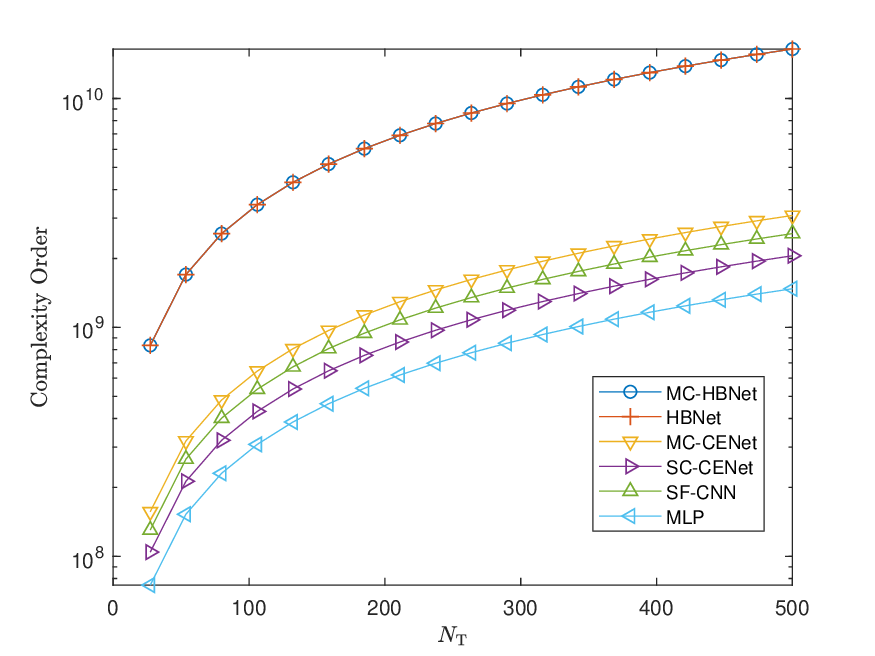} } 
		\caption{\color{black}Computational complexity order with respect to $N_\mathrm{T}$.  }
		\label{fig_complexity}
	\end{figure}
	We assessed the training times of all DLHB frameworks. We select the same simulations settings presented in Section~\ref{sec:HD_Design}. For $M=16$, Table~\ref{tableComp_Networks} lists training times for each network (Fig.~\ref{fig_Networks}) and DLHB framework (Fig.~\ref{fig_DLFrameworks}), respectively. The simple structure and smaller input/output layer sizes of MC-HBNet, HBNet, and MLP implies that they have the lowest training times than the CENet. Similarly, F1 is the fastest in training while F3 is the slowest. Note that we trained each SC-CENet separately one after the other. The training time of F3 is reduced when all SC-CENet networks are to be trained jointly in parallel. Designing hybrid beamformers by solving (\ref{PrecoderAllCarriers}) and (\ref{CombinerOnlyProblemAllSubcarriers}) using the MO algorithm introduces computational overhead. While this process is tedious, our proposed DLHB holds up this complexity only during the training. In the prediction stage, however, DLHB exhibits far smaller computational times than other algorithms.

	For the sake of completeness, Tables~\ref{tableComp_Networks2} lists the prediction stage computational times of the networks and frameworks, respectively. 	All networks show similar run times because of parallel processing of deep networks with GPUs. Among the DLHB frameworks, F1 is the fastest due to its structural simplicity. {\color{black}The MO algorithm takes longest to run in solving its inherent optimization problem. The combination of the analytical approaches, such as MO with ADCE, leads the longest computation times as compared to the other DL-based techniques even if they are used with DL-based techniques in hybrid architecture, such as ADCE+HBNet.} While GS-HB and PE-HB are quicker than F3, they are fed with the true channel matrix and lack any channel estimation stage. The F2 has slightly less execution times than GS-HB and PE-HB and provides more robust performance without requiring the CSI. Hence, we conclude that the proposed DL frameworks are computationally efficient and more tolerant to many different corruptions in the input data. 
	
	Finally, we examine the time complexity of the proposed online training approach. Suppose that $T_\mathrm{AA}$ and $T_\mathrm{DL}$ denote the time spent to estimate channel and hybrid beamformers for an analytical approach (e.g., ADCE and PE-HB), and proposed DL approach, respectively. We also define $T_\mathrm{OT}$ as the online training time, as illustrated in Table~\ref{tableOL}. {\color{black}From the previous simulations, the online algorithm requires model update about every $6^\circ$. Let us assume that a user is located $100$ m far from the BS and it moves at the circumference of the BS for $6^\circ$ at the speed of $20$ km/h. Thus, the user moves about $2$ s, which includes $200$ blocks LTE (long term evaluation) channel coherence times of $10$ ms, in which the channel statistics are not changed.   Now, let us consider a reasonable scenario of incoming data blocks for $1000$ channel coherence time, and assume that the DL network performs parameter update for every $200$ data blocks.  In this case, the time complexity of the analytical approach is $25$ s, whereas the proposed DL approach only takes $11.5$ s and $14.06$ s for offline and online settings, respectively. This shows the superiority of the proposed method for both online and offline scenario. Notice that the time complexity of online approach is slightly higher than that of the offline case due to the involvement of parameter update. Nevertheless, online training provides much better performance as shown in Fig.~\ref{fig_Online0_20}. }
	
	{\color{black} Fig.~\ref{fig_complexity} compares the complexity order of the DL networks according to Section~\ref{sec:CComp},		together with SF-CNN \cite[Table I]{deepCNN_ChannelEstimation} and MLP \cite[Table I]{mimoDeepPrecoderDesign}. The complexity of MC-HBNet and HBNet are the same and higher than the other network architectures because they include large number of units in the fully connected layers.	The MLP has the lowest complexity but it suffers from poorer feature extraction.
	}
	
	\section{Summary}
	\label{sec:Conc}
	We introduced three DL frameworks for channel estimation and hybrid beamformer design in wideband mm-Wave massive MIMO systems. Unlike prior works, the proposed DL frameworks do not require the knowledge of the perfect CSI to design the hybrid beamformers.	We investigated the performance of DLHB approaches through several numerical simulations and demonstrated that they provide {\color{black}higher spectral efficiency, and greater tolerance to corrupted channel data than the state-of-the-art, giving insights to the design of DL algorithms for wireless communications.} The robust performance results from training the deep networks for several different channel scenarios which are also corrupted by synthetic noise. This aspect has been ignored in earlier works. {\color{black}In addition, we showed that the proposed DL approach can work effectively with much less number of pilots as compared to the state-of-the-art algorithms.} We showed that the trained networks provide robust hybrid beamforming even when the received paths change up to 4 degrees from the training channel data. This allows for sufficiently long times in deep network operations without requiring re-training. This significant improvement addresses the common problem of short coherence times in a mm-Wave system. Even in terms of the channel estimation accuracy, our DLHB frameworks outperform other DL-based approaches such as SF-CNN. Our experiments show that the channel estimation performance of all DL methods maxes out at high SNR$_{\overline{\mathbf{N}}}$ regimes. This is explained by the nature of deep networks which are biased estimators. {\color{black}The label-free techniques, such as unsupervised and reinforcement learning, have the potential to reduce the complexity of the labeling process.}

	\section*{Acknowledgments}
	K. V. M. acknowledges Prof. Robert W. Heath Jr. of The University of Texas at Austin for helpful discussions and suggestions.
	
	\balance
	\bibliographystyle{IEEEtran}
	\bibliography{IEEEabrv,references_047_journal}

	\begin{IEEEbiographynophoto} {Ahmet M. Elbir} (IEEE Senior Member) received the Ph.D. degree from Middle East Technical University in 2016 He is a Senior Researcher at Duzce University, Duzce, Turkey, and Research Fellow at the Interdisciplinary Centre for Security,		Reliability and Trust (SnT), University of Luxembourg.
	\end{IEEEbiographynophoto}
	\begin{IEEEbiographynophoto} {Kumar Vijay Mishra} (IEEE Senior Member)  received his Ph.D. degree in electrical engineering from the University of Iowa. He is a U.S. National Academies Harry Diamond Distinguished Fellow at the United States Army Research
		Laboratory, Adelphi, Maryland, 20783,	USA; and an honorary research fellow at 	SnT, the University of Luxembourg.
	\end{IEEEbiographynophoto}

	\begin{IEEEbiographynophoto} {M. R. Bhavani Shankar} (IEEE Senior Member) received the  Ph.D. degree	in electrical communication engineering from the
		Indian Institute of Science, Bangalore, India, in  2007. He is currently a Research Scientist at SnT, University of Luxembourg.
	\end{IEEEbiographynophoto}
	\begin{IEEEbiographynophoto} {Bj\"orn Ottersten} (IEEE Fellow) received the Ph.D. degree in electrical engineering from 		Stanford University, Stanford, CA, USA, in 1990. He is currently the Director with the SnT, University of Luxembourg.
	\end{IEEEbiographynophoto}

\end{document}